\documentclass[%
 reprint,
 amsmath,amssymb,
 aps,
floatfix,
]{revtex4-2}

\usepackage{graphicx}
\usepackage{dcolumn}
\usepackage{bm}
\usepackage{hyperref}
\usepackage[mathlines]{lineno}
\usepackage{color}

\begin{document}

\preprint{APS/123-QED}

\title{Out-of-equilibrium quantum thermochemical \\  engine with one-dimensional Bose gas}

\author{Vijit V. Nautiyal$^{1,2}$}
 \altaffiliation[]{vijitvin94@gmail.com}
\affiliation{%
 $^{1}$School of Mathematics and Physics, University of Queensland, Brisbane, Queensland 4072, Australia. \\
$^{2}$ School of Education, The University of New England, Armidale, NSW 2350}%





\begin{abstract}
We theoretically investigate the finite-time performance of a quantum thermochemical engine utilizing a harmonically trapped one-dimensional (1D) Bose gas in the quasicondensate regime as the working fluid. The proposed engine operates in an Otto cycle, where the unitary work strokes are simulated through a quench of the interatomic interactions of 1D Bose gas. In the work strokes, the working fluid is treated as a closed quantum many-body system that undergoes dynamic evolution, beginning from an initial thermal equilibrium state at a non-zero temperature. On the other hand, during the thermalization strokes, the working fluid is treated as an open many-body quantum system in diffusive contact with a thermal reservoir, allowing particle exchange alongside the transfer of heat. Using a $c$-field approach, we demonstrate that the engine's operation is enabled by the chemical work done on the working fluid through the flow of particles from the hot reservoir. We examine the finite-time performance of the proposed quantum thermochemical engine in two extreme regimes: (\emph{i}) the \textit{ out-of-equilibrium } regime (sudden quench), which yields near-maximum power (due to fast driving of the system) while compromising efficiency, and (\emph{ii}) the \textit{quasistatic} (near-adiabatic) limit, which approaches maximum efficiency but generates zero power output due to slow driving of the system. Notably, we show that using chemical work allows the engine to achieve efficiencies close to the near-maximum (quasistatic) limit, even in the out-of-equilibrium regime, while maintaining high power output. Thus, in the out-of-equilibrium operational regime, our proposed engine provides a favourable trade-off between efficiency and power output. We also draw connections to previous research, particularly the case of an adiabatic engine cycle operating at zero temperature. We show that this zero-temperature scenario establishes an upper bound on the efficiency and work of our proposed thermochemical engine operating at non-zero temperatures. 
\begin{description}
\item[Keywords]
Quantum thermodynamics, quantum engines, Otto cycle, many-body dynamics, \\ ultracold atoms, quasicondensate, out-of-equilibrium dynamics, quantum thermochemical engines
\end{description}
\end{abstract}

\maketitle


\section{\label{sec:Introduction} Introduction}

Quantum engines (QEs) \cite{bouton2021quantum, koch2022making, myers2020bosons,nautiyal2024finite, fogarty2020many, chen2019interaction, simmons2023thermodynamic,boubakour2023interaction, keller2020feshbach, li2018efficient, li2021shortcut, singh2020optimal, shastri2022optimization, Gashu, sajitha2024quantumthermalmachineregimes, pozas2018quantum, singh2023thermodynamic, singh2022unified,singh2020performance,singh2019three, ptaszynski2022non, pedram2023quantum, williamson2023many} offer an ideal platform for testing the fundamental laws of thermodynamics in the quantum regime. One-dimensional (1D) Bose gases are a promising candidate for use as an interacting many-body working fluid in QEs \cite{chen2019interaction, li2018efficient, fogarty2020many, keller2020feshbach, nautiyal2024finite, nautiyal2024classical, chen2018bose, estrada2024quantum, charalambous2019heat, boubakour2023interaction}. The experimental accessibility and precise control over extreme physical parameter regimes in a 1D Bose gas of ultracold atoms makes it suitable for performing quantum simulations in experimentally realisable scenarios \cite{bloch2012quantum, cornish2024quantum, gross2017quantum}. In an interacting 1D Bose gas, exotic many-body quantum effects such as quantum correlations \cite{carollo2020nonequilibrium, xiao2023quantum}, many-body interactions \cite{chen2019interaction, jaramillo2016quantum, nautiyal2024finite, keller2020feshbach, boubakour2023interaction}, quantum criticality \cite{fogarty2020many}, many-body localisation \cite{halpern2019quantum} and quantum statistics \cite{koch2022making} can be exploited to facilitate operation as a QE and enhance the performance of such engines.

QEs with interacting many-body systems as the working fluid can operate in multiple ways that are not typically possible with QEs with non-interacting working fluids. For example, quenching the interaction strength can enable work to be performed on or extracted from the working fluid, similar to the compression and expansion work strokes in a traditional Otto engine cycle \cite{chen2019interaction, nautiyal2024finite, keller2020feshbach, li2018efficient}. Further, unlike a conventional Otto cycle, which relies on coupling with a thermal reservoir to increase the system's internal energy, an interacting quantum system can achieve this by quenching the interaction strength. This process alters the system's internal energy independently, without necessitating heat exchange with a reservoir. This increase in internal energy can then be harnessed to extract work from the system, as demonstrated in recent experimental studies \cite{simmons2023thermodynamic, koch2022making}.

A trade-off exists between the power output and efficiency of a QE when operated in finite-time \cite{campbell2017trade, campisi2016power, shiraishi2016universal, singh2022unified, kaur2024performance, deng2013boosting}. Generally, as the engine's driving time increases, irreversible work is produced due to non-adiabatic excitations, leading to a reduction in efficiency \cite{keller2020feshbach, shiraishi2016universal, abah2019shortcut,chand2021finite}. Conversely, irreversible work is minimized when the QE operates in the adiabatic (quasi-static) limit, allowing the engine to reach near-maximum efficiency. However, the infinitely long driving time in this scenario results in zero power output \cite{li2018efficient, abah2019shortcut, mukherjee2021many}. A central challenge in quantum thermodynamics is designing QEs that optimize the trade-off between efficiency and power output. The aim is to operate with near-maximum (near-adiabatic) efficiency while maintaining non-zero power output in finite-time engine operation \cite{andresen2011current,kaur2024performance, nautiyal2024finite, keller2020feshbach, li2018efficient, raz2016geometric, chen2024optimal, shastri2022optimization}. 

In recent studies on QEs using interacting Bose gases, efforts to enhance finite-time performance have primarily focused on implementing shortcut to adiabaticity (STA) protocols \cite{guery2019shortcuts, abah2019shortcut, keller2020feshbach, li2018efficient, fogarty2020many, campbell2017trade}. While promising, STA protocols face notable limitations, including challenges with modulation instability \cite{keller2020feshbach, li2018efficient}, increased energy requirements for practical deployment \cite{calzetta2018not}, and significant experimental constraints \cite{chen2010transient}. Consequently, exploring alternative, simpler methods to achieve near-maximum efficiency while maintaining non-zero power output has become increasingly valuable (e.g., see \cite{revathy2024improving}).

QEs operating in \textit{out-of-equilibrium} (sudden quench) regimes \cite{wiedmann2020non, kaur2024performance, nautiyal2024finite, koyanagi2024classical, pezzutto2019out, singh2020performance, denzler2024nonequilibrium, watson2024interaction,arezzo2024many, paulino2023nonequilibrium, paneru2018optimal, carollo2020nonequilibrium} have garnered significant attention in recent years for three main reasons. First, they are experimentally realizable with precise control over external parameters \cite{carollo2020nonequilibrium} and hence offer a realistic evaluation of the engine performance by accounting for non-adiabatic effects during the finite-time driving of the work strokes \cite{kaur2024performance, singh2020performance,nautiyal2024finite,carollo2020nonequilibrium}. Second, work strokes are implemented in such engines in the shortest time, providing an opportunity to maximize the power output \cite{nautiyal2024finite, arezzo2024many, bhandari2022continuous,paneru2018optimal}. Third, studies of non-equilibrium QEs can be extended to include non-equilibrium dynamics resulting from strong coupling between the working fluid and the reservoir during the thermalization strokes \cite{nautiyal2024finite, pezzutto2019out,newman2020quantum, kaneyasu2023quantum}. 

The impact of these non-equilibrium dynamics on engine performance is model-dependent and can either be advantageous or detrimental. For example, in reference \cite{newman2020quantum}, it was shown that strong coupling between the working fluid and the reservoir in non-equilibrium QEs induces system-reservoir correlations that reduce the efficiency and power output. Conversely, references \cite{wiedmann2020non, kaneyasu2023quantum} demonstrate that QEs operating in out-of-equilibrium regimes can harness quantum correlations between the working fluid and reservoirs to improve performance under certain conditions.

Even though non-equilibrium QEs, in general, have attracted attention, only a handful of studies have examined the performance of an \textit{out-of-equilibrium} (sudden quench) QE with interacting 1D Bose gas as the working fluid. Only two recent studies have provided preliminary insights to the best of our knowledge \cite{watson2024interaction, nautiyal2024finite}. This gap underscores the need to explore such systems further, especially to understand how efficiency and power trade-offs manifest under non-equilibrium conditions and determine the upper bound on the performance of such engines in realistic scenarios.

\begin{figure}[t!]
  \centering
  \includegraphics[scale=0.27]{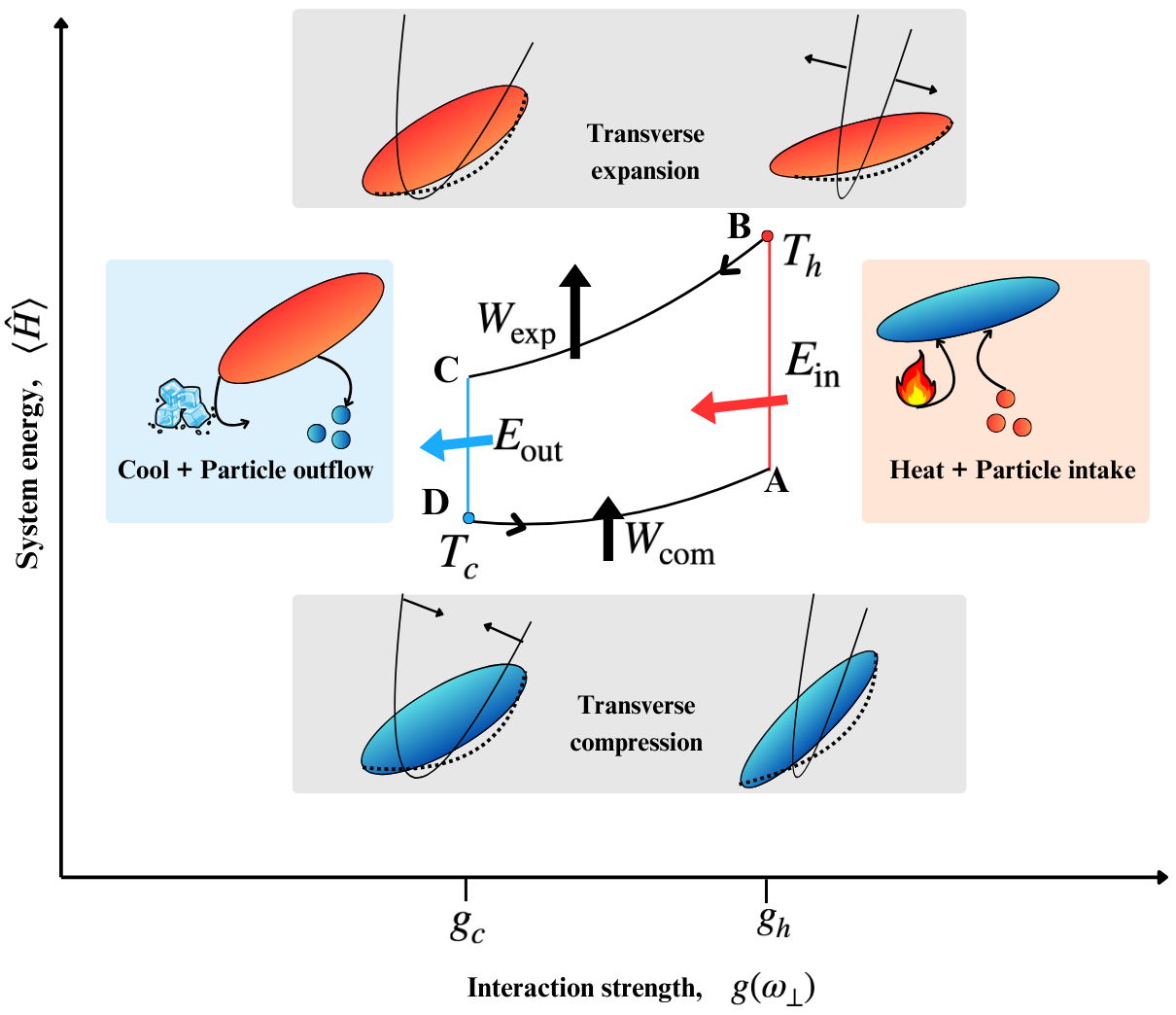}
\caption{Schematic diagram of the proposed quantum thermochemical Otto engine cycle driven by inter-atomic interactions in a weakly interacting harmonically trapped 1D Bose gas. The Otto cycle operates between two interaction strengths denoted as $g_c$ and $g_h$. The unitary work strokes, labelled \textbf{BC} and \textbf{DA}, are represented in grey. The diffusive thermalization strokes, \textbf{AB} and \textbf{CD}, are colour-coded to reflect their connection to the cold (blue) and hot (red) reservoirs, corresponding to temperatures $T_c$ and $T_h$, respectively. Since $g \simeq 2\hbar \omega_{\perp}a$, modifying the transverse confinement frequency $\omega_{\perp}$ enables the tuning of $g$. By increasing or decreasing $\omega_{\perp}$, the work strokes \textbf{DA} and \textbf{BC} become analogous to the  (transverse) compression and expansion strokes of a traditional volumetric Otto cycle. }
  \label{fig:Fig1}
\end{figure}

This work investigates the finite-time performance of an experimentally realizable quantum thermochemical engine (QTE)  operating in a non-equilibrium regime. The proposed QTE operates an Otto cycle, where the expansion and compression strokes are implemented through a sudden quench of inter-particle interactions in a weakly interacting 1D Bose gas in the quasicondensate  \cite{bouchoule2016finite,kheruntsyan2005finite} regime. During the thermalization strokes, the working fluid is in diffusive contact with a thermal reservoir, allowing chemical work in the form of particle flow in addition to heat energy. We demonstrate that the operation as an engine is enabled by this chemical work, where additional particles from the hot reservoir are introduced into the system.

Moreover, we benchmark the performance of the out-of-equilibrium (sudden quench) engine with that of a quasistatic (near-adiabatic) engine, which achieves near-maximum efficiency. Additionally, we extend the results of reference \cite{keller2020feshbach} to the finite-temperature regime of 1D Bose gas. We relate our findings to prior research in reference \cite{keller2020feshbach}, which analyzed the performance of a similar interaction-driven chemical engine in the adiabatic (maximum efficiency) regime with a zero-temperature Bose gas. We show that these earlier results at zero temperature explored in reference \cite{keller2020feshbach} provide an upper bound on the efficiency of our proposed QTE when operating at finite (non-zero) temperatures. Finally, we show that by performing thermochemical work—specifically, increasing the number of particles exchanged with the thermal reservoir— our proposed QTE can achieve \textit{near-maximum} efficiency (i.e. efficiency close to the \textit{quasistatic} limit ) even in the sudden quench regime without the need for optimization protocols such as shortcuts to adiabaticity, as employed in reference \cite{keller2020feshbach}.

The choice of a 1D Bose gas as the working fluid for our proposed QTE is motivated by three main factors. First, 1D Bose gases are routinely realised experimentally using ultracold atomic gases confined in highly anisotropic trapping potentials, which can be realized through atom chips \cite{Esteve2006,hofferberth2007non,Karen_Yang_2008,Bouchoule2011} or two-dimensional optical lattices \cite{Greiner2001,Moritz2003,Tolra2004,Kinoshita2004,Kinoshita2005}. In these experimental set-ups with anisotropic potentials, the transverse excitation energy is significantly larger than all other energy scales. Consequently, the dynamics are suppressed in the two transverse dimensions and confined to the remaining longitudinal dimension \cite{Kruger2010,Armijo2011,Kruger2022}. Further, the inter-atomic interaction strengths of the 1D Bose gases can either be tuned using confinement-induced resonances \cite{Olshanii1998,Haller2009,Haller2010} or magnetic Feshbach resonances \cite{Chin2010}, making them highly controllable in an experimental scenario.  Second, in the case of a uniform system, the ultracold 1D Bose gases can be effectively modeled through elastic two-body interactions, primarily governed by (low-energy) $s$-wave scattering processes \cite{Olshanii1998}. This implies that 1D Bose gases can be mathematically described using the integrable 1D Lieb-Liniger model \cite{liebliniger}, which, in the uniform limit, provides exact many-body solutions for the system's equilibrium thermodynamic properties via the thermodynamic Bethe ansatz. \cite{yang1969thermodynamics,kheruntsyan2003pair,Karen_Yang_2008,kerr2024analytic}. These exact solutions serve as benchmarks for approximate analytical methods and many-body computational techniques (see, e.g., \cite{kheruntsyan2003pair,Drummond2004,kerr2024analytic,Watson_Maxwell}). Third, the computational $c$-field approach based on the (stochastic) projected Gross-Pitaevskii equation (SPGPE) \cite{Blakie_cfield_2008, blakie2008dynamics, bayocboc2022dynamics, thomas2021thermalization, bayocboc2023frequency, simmons2020quantum} utilized in this work, is particularly well-suited for 1D systems. Simulating the engine cycle in 1D allows for a more detailed exploration of parameter regimes with significantly reduced computational complexity compared to the two- and three- dimensional systems. Nonetheless, the QTE proposed here, can, in principle, be realised to higher-dimensional Bose gases \cite{Blakie_cfield_2008, blakie2008dynamics}.

The structure of this article is as follows: Section \ref{sec:2} introduces the model of the Otto cycle and outlines the $c$-field approach used to simulate the finite-time engine cycle. Section \ref{sec:3} presents the main results of the paper, where we evaluate the performance of the QTE  and quantify its figures of merit. We evaluate key metrics such as power output, efficiency, and the power-efficiency trade-off as the duration of the work strokes shifts from the \textit{out-of-equilibrium} (sudden quench) regime to the \textit{quasistatic} (near-adiabatic) regime. Finally, Section \ref{sec:conclusion} provides a summary of our results and discusses potential future directions.

\section{The thermochemical Otto engine cycle} \label{sec:2}

The Otto engine cycle, illustrated in Fig.\ref{fig:Fig1}, is frequently studied in quantum thermodynamics due to its simplistic design \cite{Myers2022quantum} and close resemblance to real-world engine cycles \cite{schroeder2020introduction}. The Otto cycle consists of two unitary work strokes alternating with two isochoric thermalization strokes, where the working fluid interacts with external thermal reservoirs. Specifically, the unitary work strokes, labeled as \textbf{BC} and \textbf{DA} in Fig.\ref{fig:Fig1}, represent volumetric compression and expansion, respectively, and are executed by tuning an external parameter acting on the working fluid. The thermalization strokes, \textbf{AB} and \textbf{CD}, involve coupling the working fluid to hot and cold thermal reservoirs at temperatures $T_h$ and $T_c$, respectively, while keeping the volume constant. The following subsections will present our model and detail the numerical method for the finite-time operation of the QTE.

\subsection{The Model}
We consider a harmonically trapped 1D Bose gas of ultracold atoms in the weakly interacting quasicondensate regime \cite{Petrov_2000_Regimes, Mora-Castin-2003,kheruntsyan2005finite, kheruntsyan2003pair, garrett2013condensation, clade2009observation, Jacqmin_2011_subpoissonian} as the working fluid. The system's Hamiltonian, expressed in the second quantization formulation, is 
\begin{equation}\label{eq:hamiltonian}
     \hat{H} = \int dx \ \hat{\Psi}^{\dag} \Biggl[- \frac{\hbar^2}{2m} \frac{\partial^2}{\partial x^2} \ + \frac{1}{2} m \omega^2 x^2 + \frac{g}{2} \hat{\Psi}^{\dag} \hat{\Psi}    \Biggr] \hat{\Psi}.
\end{equation}
Here, $\omega$ represents the longitudinal trapping frequency and $m$ is the mass of one particle. The Bosonic field creation and annihilation operators are given by $\hat{\Psi}^\dagger(x)$ and $\hat{\Psi}(x)$, respectively. Additionally, $g$ is the strength of the repulsive interparticle interactions in the working fluid  ($g>0$). This quantity can be associated with the frequency of transverse confinement, $\omega_\perp$,  and the three-dimensional $s$- wave scattering length, $a$, through the relation $ g \!\simeq\! 2 \hbar \omega_{\perp}a$, which holds true away from confinement-induced resonances\cite{Olshanii1998}.

Simulating the work strokes of the Otto cycle involves modeling the unitary, real-time dynamics of the working fluid as described by Hamiltonian~(\ref{eq:hamiltonian}) in response to alterations in an external parameter. In this study, we investigate an interaction-driven Otto cycle  \cite{chen2019interaction, watson2024interaction, nautiyal2024finite, keller2020feshbach, li2018efficient} facilitated by controlling the interaction strength $g$. In practice, the interaction strength can be adjusted either by modifying the scattering length $a$
a through magnetic Feshbach resonance \cite{Chin2010} or by altering the frequency of the transverse confinement, $\omega_\perp$ \cite{schemmer2018monitoring, nautiyal2024finite, watson2024interaction}, both methods yield equivalent results as discussed herein. By increasing or decreasing $\omega_\perp$, we can view this tuning of interaction strength as transverse compression or expansion of the working fluid, respectively \cite{nautiyal2024finite, watson2024interaction}. Hence, drawing a parallel to the traditional volumetric Otto cycle, even when the interaction strength is manipulated via magnetic Feshbach resonance.

 Work $W_\mathrm{com}>0$ is performed on the working fluid in the compression stroke \textbf{DA} by increasing the interaction strength. Work  $W_\mathrm{exp}<0$ is done by the working fluid as we implement the expansion stroke \textbf{BC} through decreasing the interaction strength. The net work extracted in one complete cycle is thus $W = W_\mathrm{{com}} + W_\mathrm{{exp}}$, which must be negative, $W<0$, for the cycle to operate as an engine.

\subsection{Numerical method}
\label{subsec:numericmethod}

To simulate the finite-time operation of the quantum Otto cycle, we employ the numerical $c$-field method \cite{Blakie_cfield_2008}, which has been widely used for investigating the finite-temperature dynamics of Bose gases \cite{Blakie_cfield_2008, bayocboc2022dynamics, thomas2021thermalization, bayocboc2023frequency, simmons2020quantum, blakie2008dynamics, rooney2012stochastic, blakie2005projected, rooney2010decay}. This method involves separating the quantum field operator $\hat{\Psi}_i(x,t)$ into two regions: the $c$-field region, representing highly occupied low-energy modes that are treated as a complex field amplitude $\psi^{(\mathbf{C})}_{i}(x,t)$, and a thermal region, which contains sparsely occupied high-energy modes acting as an effective thermal bath for the $c$-field. In what follows, we describe the numerical procedure for implementing the finite-time Otto cycle using the $c$-field approach.

\subsubsection{\textbf{Stroke 1} (D $\rightarrow$ A), unitary compression work stroke} \hfill\\ We initialize the working fluid's finite-temperature thermal equilibrium state at point \textbf{D} in Fig.~\ref{fig:Fig1} by utilizing the stochastic projected Gross-Pitaevskii equation (SPGPE) \cite{gardiner2003stochastic,rooney2012stochastic},

\begin{equation}
\begin{split}
    d\psi^{(\mathbf{C})}(x,t) &= \mathcal{P}^{(\mathbf{C})} \Biggl\{-\frac{i}{\hbar} \mathcal{L}^{(\mathbf{C})} \psi^{(\mathbf{C})}(x,t) dt \nonumber \\
    &\quad + \frac{\Gamma}{k_{B}T_{c}}(\mu_{c} - \mathcal{L}^{(\mathbf{C})})\psi^{(\mathbf{C})}(x,t) dt \nonumber \\
    &\quad + dW_{\Gamma}(x,t) \Biggr\},
    \label{eq:SPGPE}
    \end{split}
    \end{equation}

where $\psi^{(\mathbf{C})}(x,t)$ represents the classical field of the system, which serves as the working fluid. In Eq.~ (\ref{eq:SPGPE}), $\mu_c$ and $T_c$ correspond to the chemical potential and temperature of the thermal region. The chemical potential $\mu_c$ sets the particle number $N_c$ within the $c$-field region (or the working fluid), $\psi^{(\mathbf{C})}(x,t)$. The operator $\mathcal{P}^{(\mathbf{C})}$ acts as a projection operator, establishing the boundary between the $c$-field and the thermal region, defined by a cut-off energy $\epsilon_\mathrm{cut}$.  $\mathcal{L}^{(\mathbf{C})}_s$ is the Gross-Pitaevskii operator and is given by
\begin{equation}
\label{eqn:GPE_operator}
    \mathcal{L}^{(\mathbf{C})} = - \frac{\hbar^2}{2m} \frac{\partial^2}{\partial x^2} + \frac{1}{2}m \omega^2 x^2 + g_c |\psi^{(\mathbf{C})}(x,t)|^2,
\end{equation}
where $ g_c$ and $T_c$ are the initial interaction strength and temperature of the working fluid at cold equilibration point \textbf{D} in the Otto cycle (Fig~\ref{fig:Fig1}). Further, the chemical potential $\mu_c$, will fix the working fluid's initial particle number,  $N_c$.

Lastly,  $dW_{\Gamma}(x,t)$ is the stochastic noise term in Eq.~(\ref{eq:SPGPE}) that is associated  with complex white noise, defined by the correlation \cite{blakie2008dynamics, gardiner2003stochastic, rooney2012stochastic},
\begin{equation}
\label{eqn:noise-correlation}
    \langle dW^*_{\Gamma}(x,t) dW_{\Gamma}(x',t) \rangle  = 2\Gamma \delta(x-x')dt.
\end{equation}
 Here, $\Gamma$ represents the growth rate, quantifying the influence of diffusive contact between the $c$-field and the effective thermal reservoir. The notation $\langle \cdot \rangle$ indicates stochastic averaging across numerous independent stochastic realizations (or trajectories). The growth rate $\Gamma$ can be selected for numerical convenience, as it does not influence the final thermal equilibrium state \cite{blakie2008dynamics, rooney2012stochastic}.

Upon completing the preparation phase, the working fluid reaches point \textbf{D} in the Otto cycle, as illustrated in Fig.~\ref{fig:Fig1}. To begin the first work stroke (compression) of the Otto cycle, labelled \textbf{DA}, we assume the working fluid is decoupled from the thermal reservoir, and its interaction strength is quenched from an initial value of $g_c$ to a final value of $g_h$. This interaction quench is carried out over a finite time using a linear quench protocol.
\begin{equation}\label{eq:g_com}
    g(t) = g_c+  \ (g_h-g_c) t / t_\mathrm{w},
\end{equation}
where $t_\mathrm{w}$ is the duration of the compression work stroke in which quench is completed. The working fluid subsequently undergoes unitary evolution, described by the projected Gross-Pitaevskii equation (PGPE) \cite{blakie2005projected,Blakie_cfield_2008}, as follows:

\begin{equation}
\begin{split}
    i \hbar \frac{\partial}{\partial t} \psi^{(\mathbf{C})}(x,t) &= \mathcal{P}^{(\mathbf{C})} \Biggl\{-\frac{\hbar^2}{2m} \frac{\partial^2}{\partial x^2} \\
    &\quad + \frac{1}{2} m \omega^2 x^2 + g(t)|\psi_\mathrm{s}^{(\mathbf{C})}|^2 \Biggr\}.
\end{split}
\label{eqn:PGPE}
\end{equation}

During this stroke, positive mechanical work ($W_\mathrm{com} > 0$) is performed on the working fluid, as we quench interaction strength from initial value $g_c$ to the final value $g_h$ in time $t_w$. The value of $W_\mathrm{com}$ is determined numerically by calculating the difference in the Hamiltonian energy of the working fluid at the culmination of the compression stroke at point \textbf{A}, specifically, $W_\mathrm{com}=\langle \hat{H}\rangle_{\mathbf{A}}-\langle\hat{H} \rangle_{\mathbf{D}}$.

The duration of the compression stroke \textbf{DA}, denoted as $t_\mathrm{w}$, dictates the state of the working fluid at the end of the stroke. If the compression is performed through a sudden quench, the fluid will reach point \textbf{A} in a highly non-equilibrium state with no well-defined temperature. On the other hand, a slow, quasistatic quench will result in the working fluid at point \textbf{A} having a well-defined temperature, $T > T_c$.

\subsubsection{\textbf{Stroke 2} (A $\rightarrow$ B ), isochoric thermalization with the hot reservoir} \label{stroke2}\hfill\\
After completion of compression work stroke \textbf{DA,} working fluid is allowed to thermalize with a hot thermal reservoir. This hot isochoric stroke is modelled using the SPGPE

\begin{equation}
\label{eq:spgpe_hot}
\begin{split}
    d\psi^{(\mathbf{C})}(x,t) &= \mathcal{P}^{(\mathbf{C})} \Biggl\{-\frac{i}{\hbar} \mathcal{L}^{(\mathbf{C})} \psi^{(\mathbf{C})}(x,t) dt \\
    &\quad + \frac{\Gamma}{k_{B}T_{h}}(\mu_{h} - \mathcal{L}^{(\mathbf{C})})\psi^{(\mathbf{C})}(x,t) dt \\
    &\quad + dW_{\Gamma}(x,t) \Biggr\}.
\end{split}
\end{equation}

Here, $T_h$ represents the temperature of the hot reservoir, and $\mu_h$ is its chemical potential. In this scenario, the bath's chemical potential, $\mu_h$, determines the particle number in the working fluid,$N_h$ at the end of the thermalization stroke \textbf{AB}. We can set $\mu_h$ so that there is no net change in particle number by the end of stroke \textbf{AB}, ensuring that $N_h = N_c$. Alternatively, $\mu_h$ can be adjusted to allow particle exchange between the working fluid and the reservoir, resulting in $N_h = N_c + \Delta N$ at the end of the stroke. The Gross-Pitaevskii operator, $\mathcal{L}^{(\mathbf{C})}$, remains the same as in Eq.~\ref{eqn:GPE_operator}, except that the interaction strength is fixed at $g = g_h$ during this stroke. During this process, energy is transferred from the hot reservoir to the working fluid, with the amount given by $E_\mathrm{in} = \langle \hat{H} \rangle_{\mathbf{B}} - \langle \hat{H} \rangle_{\mathbf{A}} > 0$.

Particle exchange between the working fluid and the thermal bath is possible because the $c$-field theory with the SPGPE formalism considers a diffusive contact between the highly occupied $c$-field region (working fluid) and the thermal reservoir, which consists of sparsely occupied high-energy modes \cite{bradley2015low, rooney2012stochastic}. This formalism enables us to theoretically model the isochoric thermalization strokes for operating as a QTE, as it allows for additional chemical work to be performed on the working fluid through the flow of particles from the hot reservoir.

\subsubsection{\textbf{Stroke 3} (B $\rightarrow$ C), unitary expansion work stroke}\hfill\\ Following the thermalization stroke with the hot reservoir \textbf{AB}, we initiate the expansion work stroke. During this stroke, the working fluid evolves according to the PGPE described in Eq.~\ref{eqn:PGPE}, but with the interaction strength, $g(t)$, being reduced from $g_h$ to $g_c$ following a linear protocol:
\begin{equation}\label{eq:g_exp}
    g(t) = g_{h}-(g_{h}-g_{c}) t  / t_\mathrm{w}.
\end{equation}
Throughout the expansion work stroke, which lasts for the same duration, $t_\mathrm{w}$, as the compression stroke, the working fluid performs work, $W_\mathrm{exp} < 0$. Similar to the calculation of $W_\mathrm{com}$, the value of $W_\mathrm{exp}$ is determined numerically by evaluating the change in the Hamiltonian energy of the working fluid at the end of the expansion stroke, concluding at point \textbf{B}. Specifically, $W_\mathrm{exp}$ is expressed as $W_\mathrm{exp} = \langle \hat{H} \rangle_{\mathbf{C}} - \langle \hat{H} \rangle_{\mathbf{B}}<0$.

\subsubsection{\textbf{Stroke 4}(C $\rightarrow$ D), isochoric thermalization with the cold reservoir reservoir} \hfill\\
\label{subsec:stroke4}
After completing the expansion stroke \textbf{BC}, we perform the thermalization stroke \textbf{CD} with the cold reservoir using the SPGPE given in Eq.~\ref{eq:SPGPE}, bringing the system back to the starting point \textbf{D}. During this stroke, the chemical potential is adjusted from $\mu_h$ to $\mu_c$ to restore the particle number $N_c$ to its original value at the start of the cycle, i.e.  $N_c = N_h - \Delta N$. During this stroke, energy $E_\mathrm{out} = \langle \hat{H} \rangle_{\mathbf{D}}-\langle \hat{H} \rangle_{\mathbf{C}}<0$ is transferred from the working fluid to the cold reservoir, bringing the working fluid and the entire Otto cycle back to its starting point at \textbf{D}.

The overall performance of the engine cycle can be assessed by calculating the net work, \begin{equation} W = W_\mathrm{{com}} + W_\mathrm{{exp}}, \end{equation} which must be negative for the Otto cycle to function as an engine. The efficiency is defined as the ratio of beneficial net work ($-W)$ to the cost, i.e. the input energy supplied to the system ($E_\mathrm{in})$ \cite{schroeder2020introduction}. Thus, efficiency is given by \begin{equation} \eta = - W/E_\mathrm{in}, \end{equation} and the power output is expressed as 

\begin{equation} 
P = -W/2t_\mathrm{w}.
\label{eq:power}
\end{equation} 

Within the scope of this study, we do not include the timescales of thermalization strokes for calculating the power output in Eq.~(\ref{eq:power}). We assume that their duration is negligible. This assumption is reasonable if we consider the reservoir to be infinitely larger than the working fluid, causing the fluid to thermalize almost instantaneously, as suggested in previous studies \cite{fogarty2020many, li2018efficient, keller2020feshbach, watson2024interaction, boubakour2023interaction}. This assumption allows us to focus on evaluating the engine's performance based solely on the duration of the work strokes, specifically examining how the quantum engine operates when these work strokes are implemented through a sudden quench of the interaction strength \cite{schemmer2018monitoring,nautiyal2024finite, mazza2014interaction,collura2018quantum, watson2024interaction} to maximize the power output.

All four strokes of the proposed QTE can be realised with existing experimental techniques. The unitary (isolated) interaction-induced work strokes can be realized in experiments either by quenching the frequency of the transverse trapping potential, as demonstrated in Ref.\cite{schemmer2018monitoring}, or by tuning an external magnetic field about a Feshbach resonance, as shown in Refs.\cite{inouye1998observation, stenger1999strongly}. Additionally, during the thermalization strokes, the working fluid is coupled to a thermal reservoir, which, in principle, allows for both particle and heat exchange  \cite{estrada2024quantum, marzolino2024quantum}. In an experimental setting, these strokes can be implemented by treating partially condensed matter waves as the working fluid, while high-energy atoms in the thermal cloud (or atoms in sparsely occupied high-energy modes) serve as an effective thermal reservoir \cite{rooney2010decay, rooney2012stochastic, rooney2016reservoir} (see \textbf{Appendix~B} for details). This approach has been employed in previous experiments \cite{weiler2008spontaneous, rooney2013persistent}, where the experimental results were simulated using the SPGPE. Specifically, in Ref.~\cite{weiler2008spontaneous}, the authors conducted temperature quenches and measured the change in particle number over time (see Fig.~2(a) of \cite{weiler2008spontaneous}). A similar method can be applied in an experimental realization of our proposed QTE, where the temperature of the thermal cloud is quenched between the hot and cold reservoir temperatures, $T_h$ and $T_c$, while adjusting the chemical potential to achieve the desired particle number in the working fluid, thereby simulating the thermalization strokes with a reservoir.

\subsection{Operation as a QTE enabled via particle flow during thermalisation stroke with the hot reservoir}
The \textit{thermochemical} Otto engine studied here is fundamentally different from a conventional Otto \textit{heat} engine \cite{thermochemicalengine}. In a conventional \textit{heat} engine, the system-reservoir coupling follows a \textit{canonical} ensemble description, allowing only heat exchange between the working fluid and the reservoir. In contrast, a \textit{thermochemical} engine operates under a \textit{grand-canonical} ensemble \cite{marzolino2024quantum,luo2024critical,estrada2024quantum}, enabling the exchange of both heat and particles (see, for example, \cite{nautiyal2024finite,eglinton2023thermodynamic, marzolino2024quantum, luo2024critical,estrada2024quantum}).

In our proposed QTE, operation as a \textit{thermochemical} engine is enabled via doing additional chemical work on the working fluid during the thermalization stroke \textbf{AB}. Specifically, chemical work is introduced through the inflow of $\Delta N$ particles from the hot reservoir into the working fluid as described in  Sec.~\ref{stroke2}. The inflow of $\Delta N$ particles increases the total particle number of the working fluid, which, in turn, results in the corresponding rise in the internal energy of the working fluid. This increase in energy of the working fluid can then be converted into mechanical work during the subsequent expansion stroke \textbf{BC}. Once this expansion work stroke is completed, the working fluid is brought into contact with the cold reservoir, allowing the same excess number of particles, $\Delta N$, to be transferred to the cold reservoir during the thermalization stroke \textbf{CD} (as described in Sec.~\ref{subsec:stroke4}), thereby restoring the working fluid to its original state with the same initial number of particles.

The efficiency of this \textit{thermochemical} Otto engine, defined as $\eta = -W/E_\mathrm{in}$, can be determined by evaluating the energy differences at the end of each stroke, as outlined in Sec.~\ref{subsec:numericmethod}. Unlike a conventional heat engine, where the energy, $E_\mathrm{in}$, supplied to the working fluid from the hot reservoir consists solely of heat transfer (through a canonical ensemble description), the energy supplied to the working fluid in the Otto \textit{thermochemical} engine, $E_\mathrm{in} = \langle \hat{H} \rangle_{\mathbf{B}}-\langle \hat{H} \rangle_{\mathbf{A}}>0$, includes a contribution from the chemical work (in addition to heat) and can be expressed as:
\begin{equation}
\label{eq:chemplusheat}
E_\mathrm{in} = Q_h + W_\mathrm{chem}.
\end{equation}
Here, $Q_h$ represents the heat absorbed by the working fluid during the thermalization stroke \textbf{AB} with the hot reservoir, while $W_\mathrm{chem}$ accounts for the additional chemical work arising from the transfer of $\Delta N$ particles.

If we adjust the chemical potential of the thermal reservoirs such that no net exchange of particles occurs, i.e., $\Delta N \simeq 0$, then $W_\mathrm{chem} = 0$, and the net increase in the internal energy of the working fluid at the end of the thermalization stroke will be solely due to heat transfer, i.e., $E_\mathrm{in} = Q_h$. In this work, we refer to this operational scheme, where $\Delta N \simeq 0$, as a quantum heat engine (QHE). Conversely, when $\Delta N \neq 0$, the engine operates as a quantum thermochemical engine (QTE) due to the additional chemical work facilitated through particle exchange.

Determining the individual contributions of $Q_h$ and $W_{\mathrm{chem}}$ to the total energy, $E_{\mathrm{in}}$, is a nontrivial task due to the intrinsic coupling between heat and particle transport in ultracold atomic gases (see, e.g., \cite{brantut2013thermoelectric, husmann2018breakdown}). However, we emphasize that the additional chemical work is accounted for in the total energetic cost when evaluating the efficiency of our proposed QTE by defining efficiency as
$\eta = -W/E_\mathrm{in}$ rather than $\eta = -W/Q_h$.

\subsection{Timescales corresponding to the out-of-equilibrium (sudden quench) and quasistatic regime of engine operation}
\label{sec:timescale}
We explore the performance of the proposed quantum Otto engine in two extreme regimes of operation with respect to the duration of the work strokes. These regimes are: (\emph{i}) \textit{the out-of-equilibrium }(sudden quench) regime, where work strokes are completed in the shortest possible time, inducing maximum irreversible work due to non-adiabatic excitations, resulting in a minimum efficiency \cite{born1928beweis,abah2019shortcut,li2021shortcut,del2013shortcuts, rezek2006irreversible}; (\emph{ii}) \textit{the quasistatic} (near-adiabatic) regime, which allows for near-maximum efficiency at the cost of minimal power, as it requires extremely slow driving of the system during work strokes to minimize the non-adiabatic excitations \cite{born1928beweis, guery2019shortcuts, li2021shortcut}.

Theoretically, a sudden quench is often treated as an instantaneous quench \cite{bayocboc2023frequency, bouchoule2016finite, thomas2021thermalization, bayocboc2022dynamics, simmons2020quantum}. In practical implementations, however, this ``sudden quench'' occurs over a finite duration \cite{schemmer2018monitoring}. This duration should be short enough to approximate a sudden (near-instantaneous) quench relative to the characteristic timescales corresponding to the longitudinal dynamics $t_{\parallel}=2\pi/\omega$, while also being slow enough to be considered near-adiabatic with respect to the characteristic timescales for the transverse dynamics, $t_{\perp}=2\pi/\omega_{\perp}$. This prevents excitation of transverse modes and thus preserves the 1D characteristic of the system \cite{schemmer2018monitoring, Olshanii1998}.

Accordingly, we simulate the unitary work strokes for the sudden quench regime within a finite time by linearly varying the interaction strength as in Eqs.~(\ref{eq:g_com}) and (\ref{eq:g_exp}), meeting the following criterion:

\begin{equation} 
\label{suddentime}
t_{\perp} \ll t_\mathrm{w} \ll t_{\parallel}. 
\end{equation}

The sudden quench regime, in which work strokes are implemented according to Eq.~(\ref{suddentime}), is referred to as the \textit{out-of-equilibrium} regime in this work. This is because these rapid quenches drive the working fluid out of equilibrium, inducing non-equilibrium dynamics, specifically in the form of breathing modes \cite{nautiyal2024finite, bouchoule2016finite, bayocboc2023frequency, schemmer2018monitoring}, which become apparent when the system continues to evolve after the quench. These out-equilibrium dynamics have been observed experimentally in reference \cite{schemmer2018monitoring}. Since power, $P=-W/2t_\mathrm{w}$, the sudden quench engine provides an opportunity to operate at near-maximum power due to short driving time, $t_\mathrm{w}$ \cite{nautiyal2024classical,nautiyal2024finite,watson2024interaction}.

In contrast, in the \textit{quasistatic} (or near-adiabatic) regime, work strokes are performed slowly over a timescale, maintaining the system in an approximate thermal equilibrium state throughout the process. Under these conditions, no breathing modes emerge either during or after the quench, even as the system continues to evolve \cite{nautiyal2024classical, nautiyal2024finite,pezzutto2019out}. We therefore refer to this timescale as \textit{quasistatic} regime of operation. For operation in the \textit{quasistatic} regime, the duration of the work strokes approach adiabatic conditions with respect to both longitudinal and transverse timescales mentioned above, i.e.,
\begin{equation} 
\label{eq:t_ad}
 t_{\perp} \ll t_{\parallel} \ll t_{\mathrm{w}}. 
 \end{equation}
Since power, $P=-W/2t_\mathrm{w}$, the quasistatic regime produces vanishing power output, $P\rightarrow 0$ (see Fig.~\ref{fig:Fig2}(c)), due to extremely long driving time, i.e. for $t_\mathrm{w} \rightarrow \infty$ \cite{nautiyal2024finite,abah2019shortcut, pezzutto2019out}.

\section{Results and discussion} \label{sec:3}

In this section, we present the main results and discuss their implications. We begin by introducing the parameters used to benchmark the performance of the \textit{out-of-equilibrium} QTE in Sec.~\ref{parameter_benchmark}. This is followed by a detailed discussion of the key findings in subsequent sections, from section \ref{chemicalworkenabled}  to section \ref{ratioeffect}.

\subsection{Parameters for benchmarking performance of out-of-equilibrium QTE}
\label{parameter_benchmark}

The operation of the interaction-driven QTE is notably different from that of a conventional Otto heat engine. QTE has a diffusive contact with the thermal reservoirs \cite{marzolino2024quantum}, allowing for a coupled flow of heat and particles (see references \cite{brantut2013thermoelectric,husmann2018breakdown} for examples) between the many-body working fluid and the reservoirs. We used the following criteria to benchmark the performance of our proposed QTE operating in the \textit{out-of-equilibrium} regime:\\

\noindent (\emph{i}) \textit{Adiabatic QTE operating at zero temperature} \cite{keller2020feshbach}: We will compare our proposed QTE's performance against QTE with an interacting Bose gas at zero temperature as the working fluid. The adiabatic QTE, with a weakly interacting 1D Bose gas as the working fluid in a harmonic trap, was first studied in reference \cite{keller2020feshbach}. In that work, the Thomas-Fermi approximation was applied to derive a simplified expression for maximum efficiency, given by \cite{keller2020feshbach}
\begin{equation}
    \eta_\mathrm{max}(T=0) = 1 - \Bigg( \frac{g_\mathrm{c}}{g_\mathrm{h}}\Bigg)^{2/3}.
    \label{eq:max_efficiency}
\end{equation}
Similarly, net work was also  calculated using the Thomas-Fermi energies at the end points of the engine cycle (points \textbf{D} and \textbf{B} in Fig.~\ref{fig:Fig1}), and is given by
\begin{equation}
    W_\mathrm{max}(T=0) = -\frac{3}{5} \Bigg(\frac{9}{32}\Bigg)^{1/3} \Bigg(\bar{g}^{2/3}_h - \bar{g}^{2/3}_c\Bigg)\ \Bigg(N^{5/3}_h - N_c^{5/3}\Bigg).
    \label{eq:maxwork}
\end{equation}
Here, $\bar{g}_{c(h)} = g_{c(h)}/(\hbar \omega l_\mathrm{ho})$ is the dimensionless interaction strength defined in terms of characteristic energy scale, $\hbar \omega$ and length scale, $l_\mathrm{ho}= \sqrt{\frac{\hbar}{m \omega}}$.

In this work, we extend the results of reference \cite{keller2020feshbach} to finite (non-zero) temperatures using the $c$-field method, as outlined in Sec.~\ref{subsec:numericmethod}. \\

\noindent (\emph{ii}) \textit{Quasistatic finite-time QTE}: To assess the trade-off between efficiency and power output for the \textit{out-of-equilibrium} (sudden quench) engine, we will compare the efficiency drop in the sudden quench engine to the maximum achievable efficiency in finite time, which is obtained by performing the work strokes quasistatically (i.e., extremely slowly) \cite{abah2019shortcut,li2021shortcut,keller2020feshbach}. This quasistatic efficiency will serve as the upper bound for finite-time efficiency in our proposed QTE. This approach is justified, as quasistatic operation yields efficiencies that approach the maximum efficiency achievable through fully adiabatic driving of the engine (see Sec.~\ref{sec:timescale} for details) \cite{abah2019shortcut, campbell2017trade, kaur2024performance, keller2020feshbach}.\\

\noindent (\emph{iii}) \textit{Efficient power criterion}: To analyse the finite-time performance of the QTE, we use an additional metric known as ``efficient power'', originally introduced in reference \cite{yilmaz2006new}. Efficient power is defined as the product of efficiency and power output, i.e.,
\begin{equation}
    \Theta = \eta \ \times P.
\end{equation}

The efficient power, $\Theta$, is a relatively new performance criterion used to assess the trade-off between power and efficiency in real-world heat engines. It provides insight into how much power output increases per unit decrease in efficiency \cite{yilmaz2006new}. Recent studies on quantum engines have incorporated this criterion as an additional metric—alongside power and efficiency—to evaluate finite-time performance and the efficiency-power trade-off \cite{myers2020bosons, nautiyal2024finite, singh2018low}.

\begin{figure*}[t!]
    \centering
    \includegraphics[scale=0.40]{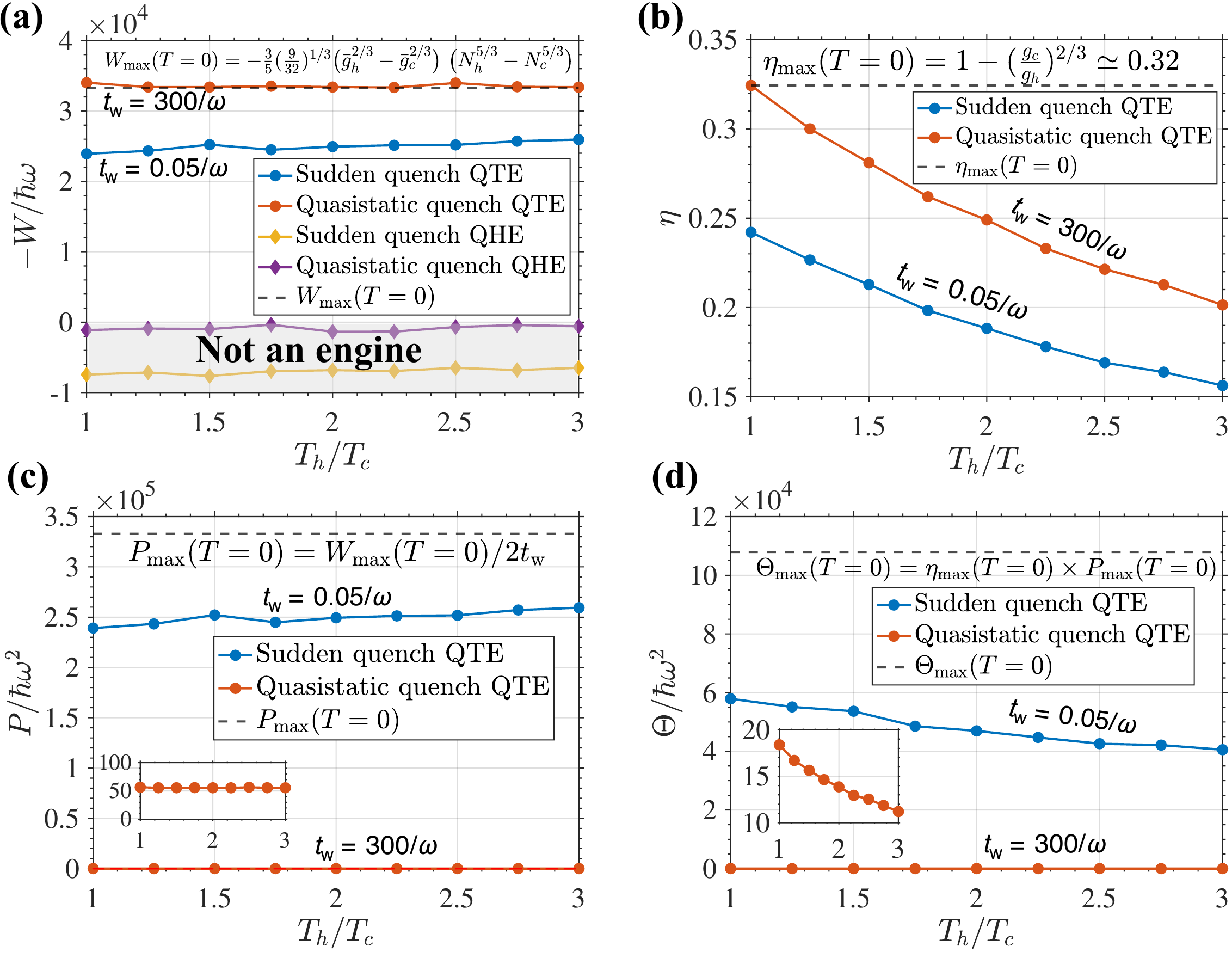}
    \caption{(a) Net work, $-W$; (b) efficiency, $\eta$; (c) power output, $P$; and (d) efficient power, $\Theta$ as a function of the temperature ratio between hot and cold reservoirs, $T_h/T_c$, for the interaction-driven quantum Otto engine with a weakly interacting 1D Bose gas. In all panels, red and blue curves with circular markers represent the operation as a quantum \textit{thermochemical} engine (QTE), where the working fluid absorbs approximately $\Delta N \simeq 1000$ particles from the hot reservoir during thermalization strokes. The yellow and purple curves (diamond markers) in panel (a) correspond to the operation as a quantum heat engine (QHE) when there is no net exchange of particles with the reservoirs, i.e., $\Delta N \simeq0$. Negative net work ($-W > 0$) is required for engine operation; the shaded region in panel (a) indicates the region where $-W < 0$, implying that operation as an engine is not possible. Data points with $-W < 0$ in panel (a) (i.e., the yellow and purple curves) are not plotted in panels (b), (c), and (d). The black dashed line in all the panels corresponds to an interaction-driven chemical engine operating at zero temperature in the adiabatic (maximum-efficiency) regime as investigated in Ref.~\cite{keller2020feshbach}. In all cases, the compression ratio was fixed at $g_h/g_c=1.8$, meaning the interaction strength was quenched from $g_c = 1.27 \times 10^{-38}$ J $\cdot$ m to $g_h=1.8 \times g_c$ following Eq.(\ref{eq:g_com}); during expansion, it was quenched back to $g_c$ according to Eq.(\ref{eq:g_exp}). The sudden quench work strokes are completed in time, $t_\mathrm{w}=0.05/\omega$, whereas the quasistatic quench work strokes are completed in time, $t_\mathrm{w}=300/\omega$. See \cite{parameters} for details on other relevant system parameters.}
    \label{fig:Fig2}
\end{figure*}

\subsection{Engine operation enabled by chemical work -  heat engine versus thermochemical engine}
\label{chemicalworkenabled}

In Fig.~\ref{fig:Fig2}(a), we compare the engine’s performance under two operating schemes: (\emph{i}) as a quantum heat engine (QHE), where no chemical work is done on the working fluid ($\Delta N \simeq 0$ and hence, $W_\mathrm{chem}=0$ in Eq.~\eqref{eq:chemplusheat}); and (\emph{ii}) as a quantum thermochemical engine (QTE), where additional chemical work is done in the form of particle flow from the hot reservoir ($\Delta N \neq 0$), along with heat. The results show that in the QHE scheme (purple and yellow curves in Fig.~\ref{fig:Fig2}(a)), the engine operation is not possible, as the net work remains positive, $W > 0$ (or $-W < 0$), across both the quasistatic and sudden quench regimes. Positive net work, $W>0$ at the end of the cycle, means that the working fluid absorbed energy during the cycle instead of converting it into useful work output. This indicates that relying purely on heat is insufficient for engine operation in the proposed Otto cycle.

The primary factor preventing the Otto cycle from operating effectively as a heat engine with lies in the dependence of net work on local atom-atom correlation and density profile of the 1D Bose gas. Rather than relying solely on the difference in local atom-atom correlation functions at the hot and cold thermal equilibrium points \textbf{B} and \textbf{D} in Fig.~\ref{fig:Fig1} (see \cite{watson2024interaction} for details), the net work also varies with the inhomogeneity of the density profile, particularly the peak density at these points. Consequently, while atom-atom pair correlations exhibit a temperature dependence favourable for producing a significant positive network, the temperature dependence of peak density works against this outcome, ultimately cancelling the positive net work that would otherwise be achievable in a uniform system with matching densities at the hot and cold equilibrium points as in references \cite{watson2024interaction} and \cite{chen2019interaction}.

Further in Fig.~\ref{fig:Fig2}(a) we observe that while it is not feasible to operate our interaction-driven Otto cycle as a QHE, we find that it can function as a quantum thermochemical engine \cite{marzolino2024quantum} enabled by additional chemical work done on the working fluid during the thermalization stroke \textbf{AB}. This extra chemical work is introduced through the inflow of particles, $\Delta N$, from the hot reservoir into the working fluid. Under these conditions, engine operation is enabled, as the increase in internal energy of the working fluid—specifically, the input energy, $E_\mathrm{in}$—is sufficiently large to produce large negative net work, i.e.,  $W<0$(or$-W > 0$ (red and blue curves) in Fig.\ref{fig:Fig2}(a), across both out-of-equilibrium (blue curve) and quasistatic (red curve) regimes.

Next, We examine the performance of \textit{out-of-equilibrium} (sudden quench) engine across various parameters of the weakly interacting 1D Bose gas, including the temperature ratio between the hot and cold reservoirs, $T_h/T_c$ (see Sec.~\ref{temeffect}) the impact of chemical work, i.e. the number of particles $\Delta N$ exchanged with the reservoirs on engine performance (see Sec.~\ref{deltaNeffect}) and the effect of the compression ratio for the work strokes, $g_h/g_c$  (see Sec.~\ref{ratioeffect}). Additionally, we will also evaluate the trade-off between efficiency and power output (see Sec.~\ref{tradeoff}) in the out-of-equilibrium engine.

\subsection{Favourable trade-off between efficiency and power output}
\label{tradeoff}

When operating in the \textit{out-of-equilibrium} (sudden quench) regime, the proposed QTE exhibits a favourable trade-off between efficiency and power output in finite-time operation. This is illustrated in Fig.~\ref{fig:Fig2}, where we plot key quantities—work, efficiency, power, and efficient power—as a function of the temperature ratio between the hot and cold reservoirs, $T_h/T_c$. We fixed the temperature of the cold reservoir, $T_c$ and evaluated quantities of interest at different values of the temperature of the hot reservoir, $T_h$.

In Fig.~\ref{fig:Fig2}(a), we compare the net work obtained in two extreme regimes of QTE operation: (\emph{i}) the sudden quench regime (blue curve), which results in minimal efficiency due to the production of maximum irreversible work from non-adiabatic excitations \cite{abah2019shortcut, del2013shortcuts}; and (\emph{ii}) the quasistatic quench regime (red curve), which produces near-maximum efficiency in finite-time operations \cite{abah2019shortcut, nautiyal2024finite}.

In the quasistatic regime, our numerical results show excellent agreement with the analytical zero-temperature results (black dashed curve). The QTE produces the maximum possible work, closely matching the theoretical maximum work given by Eq.(\ref{eq:maxwork}) for an adiabatic engine operating at zero temperature (black dashed line), as studied in Ref.\cite{keller2020feshbach}. This outcome is expected since the quasistatic quench is slow enough to avoid generating non-adiabatic excitations. Notably, however, as we transition from the quasistatic to the sudden quench regime, the reduction in net beneficial work remains relatively minor, even though the work strokes are performed orders of magnitude faster. This finding suggests that, in a system of weakly interacting 1D Bose gas, the irreversible work generated by non-adiabatic excitations is not substantial enough to significantly impact efficiency, as efficiency is given by the ratio of net work, $-W$, to input energy, $E_\mathrm{in}$. This is confirmed in the Fig.~\ref{fig:Fig2}(b), where we plot the efficiencies.

 In Fig.\ref{fig:Fig2}(b), we show the efficiency as a function of temperature for both the sudden quench QTE (blue curve) and the quasistatic QTE (red curve). The results reveal that transitioning from the sudden quench regime to the quasistatic regime does not result in a significant efficiency loss. Remarkably, even though the work strokes in the sudden quench engine are completed orders of magnitude faster, it operates at efficiencies close to the maximum efficiency achieved by the quasistatic engine. These findings are consistent with the results in Fig.~\ref{fig:Fig2}(a), where the transition from the quasistatic regime to the sudden quench regime shows minimal loss in work output. This enables the sudden quench engine to operate at efficiencies close to the maximum limit achieved by the quasistatic engine, as shown in Fig~\ref{fig:Fig2}(b).

In Fig.\ref{fig:Fig2}(a), we observe a negative impact of temperature on engine performance in both the sudden and quasistatic quench regimes. Specifically, when there is no temperature difference between the reservoirs (i.e., $T_h = T_c$), the quasistatic engine achieves an efficiency equivalent to that of the QTE operating at zero temperature ($\eta_\mathrm{max}(T=0)$), as represented by the black dashed line, and investigated in Ref.\cite{keller2020feshbach} (see Eq.(\ref{eq:max_efficiency}) for the expression of $\eta_\mathrm{max}(T=0)$). However, unlike a conventional Otto heat engine, the performance of the QTE deteriorates as the temperature of the hot reservoir increases. This effect is evident in Fig.\ref{fig:Fig2}(b), where both out-of-equilibrium and quasistatic engines display a decrease in efficiency as the reservoir temperature rises. This suggests that the additional heat supplied to the working fluid is merely raising the operational cost, i.e., $E_\mathrm{in}$, without being converted into useful work. This finding aligns with our results in Fig.\ref{fig:Fig2}(a), where we demonstrate that heat alone is insufficient for engine operation (see Sec.~\ref{chemicalworkenabled} for details) as in the absence of chemical work ($\Delta N \simeq 0$), engine operation is not feasible (yellow and purple curves).

\begin{figure}[t!]
    \includegraphics[scale=0.44]{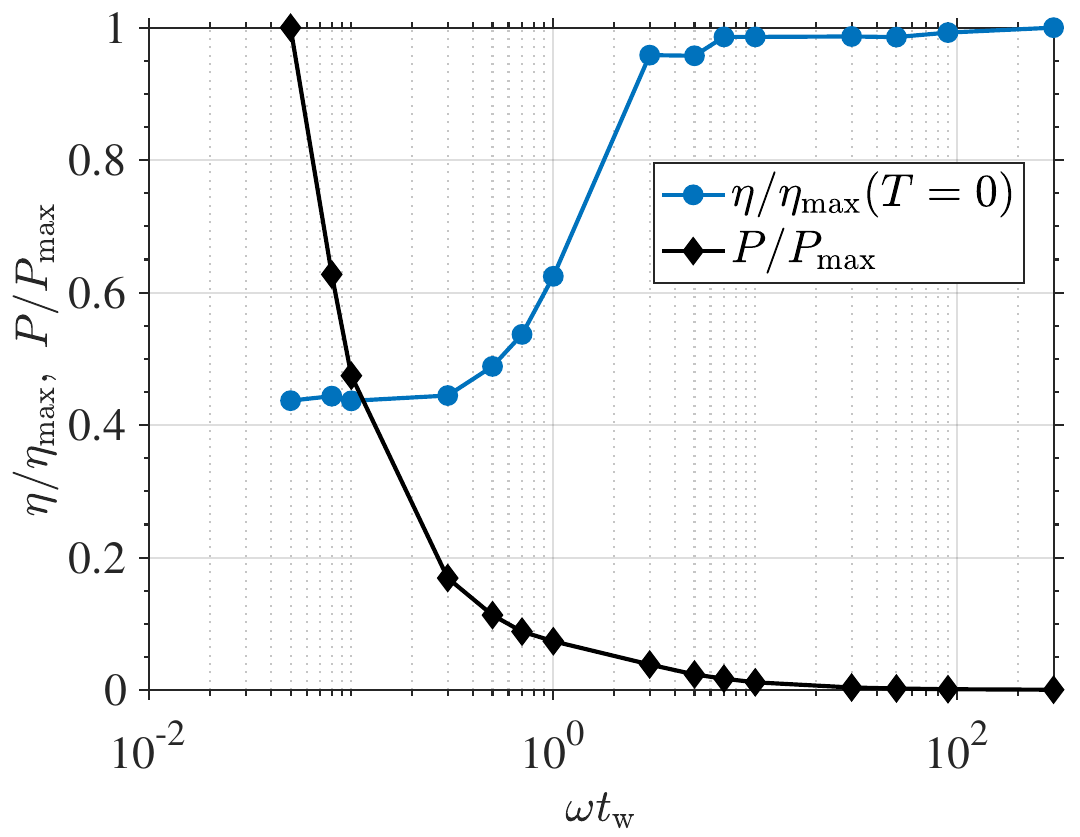}
    \caption{Efficiency, $\eta$, and power, $P$, as functions of the duration of the work strokes, $t_\mathrm{w}$. Efficiency is normalized to the maximum adiabatic efficiency of a QTE with a zero-temperature Bose gas, $\eta_\mathrm{max}(T=0) \equiv \eta_\mathrm{max}$ (see Sec.\ref{parameter_benchmark} and Eq.\ref{eq:max_efficiency}). Power output is normalized to the near-maximum finite-time power, which corresponds to the shortest timescale considered in our model, $t_\mathrm{w} = 0.05/\omega$, i.e., $P(t_\mathrm{w} = 0.05/\omega) \equiv P_\mathrm{max}$. All other parameters are the same as in Fig.~\ref{fig:Fig2}, with a temperature ratio between the hot and cold reservoirs of $T_h/T_c = 1$.} 
    \label{fig:Fig:tq}
\end{figure}

In Fig.\ref{fig:Fig2}(c) and Fig.\ref{fig:Fig2}(d), we plot power output and efficient power as a function of the temperature ratio, $T_h/T_c$, for both sudden quench and quasistatic quench engines. As expected, the quasistatic engine (red curve) produces nearly zero power and efficient power due to its extremely long cycle time. In contrast, the sudden quench engine achieves power output and efficient power values orders of magnitude higher than those of the quasistatic engine. This suggests that the efficiency loss per unit power is relatively minor in the sudden quench engine, allowing it to significantly outperform the quasistatic engine in practical finite-time operations. This advantage is particularly noteworthy as no optimization protocols, such as the STA, were applied in the sudden quench regime.

Fig.~\ref{fig:Fig:tq} illustrates the trade-off between power and efficiency in the finite-time operation of the QTE. Specifically, efficiency and power are plotted as functions of the quench duration of the work strokes, $t_\mathrm{w}$. As expected, increasing efficiency in finite-time operations comes at the cost of power output, i.e., as $\eta \rightarrow\eta_\mathrm{max}(T=0)$(see Eq.~\eqref{eq:max_efficiency}), power approaches zero, $P \rightarrow 0$. Transitioning from the sudden quench regime ($t_\mathrm{w} = 0.05/\omega$) to the quasistatic regime ($t_\mathrm{w} = 300/\omega$) increases efficiency while the power steadily declines to zero.

In our model, we observed that the highest power in finite-time operations, $P_\mathrm{max}$, is achieved at the shortest work stroke duration, $t_\mathrm{w} = 0.05/\omega$, in the sudden quench regime. At such short timescales, efficiency is reduced due to irreversible work generated by quantum friction \cite{abah2019shortcut, del2013shortcuts}. In a 1D Bose gas, this irreversible work manifests as breathing mode excitations \cite{nautiyal2024finite, bouchoule2016finite, bayocboc2023frequency, tschischik2013breathing}, arising when work strokes are executed on timescales much shorter than the longitudinal trap period, i.e., $t_\mathrm{w} \ll 2\pi/\omega$ (see Section \ref{sec:timescale}). However, even at the shortest duration of $t_\mathrm{w}=0.05/\omega$, ($P/P_\mathrm{max} \rightarrow 1$), the QTE still achieves around 43\% of the zero-temperature adiabatic efficiency limit, $\eta_\mathrm{max}(T=0)\equiv \eta_\mathrm{max}$ (see Eq.~\eqref{eq:max_efficiency}), i.e., $\eta/\eta_\mathrm{max} \simeq 0.43$ at shortest quench time, $t_\mathrm{w}=0.05/\omega$.

Further, we identify a cutoff time at approximately $t_\mathrm{w}\simeq10/\omega$ beyond which the efficiency saturates, achieving a near-maximum value and showing no significant further gains. This suggests that beyond this timescale, the work strokes are slow enough to suppress non-adiabatic excitations \cite{nautiyal2024finite, keller2020feshbach, abah2019shortcut, del2013shortcuts} that would otherwise contribute to irreversible work and reduce the engine’s efficiency. Overall, these results reinforce the findings in Fig.\ref{fig:Fig2}, confirming that the efficiency drop in the sudden quench regime is not substantial and that the trade-off between efficiency and power remains favourable.

\begin{figure*}[t!]
    \centering
    \includegraphics[scale=0.40]{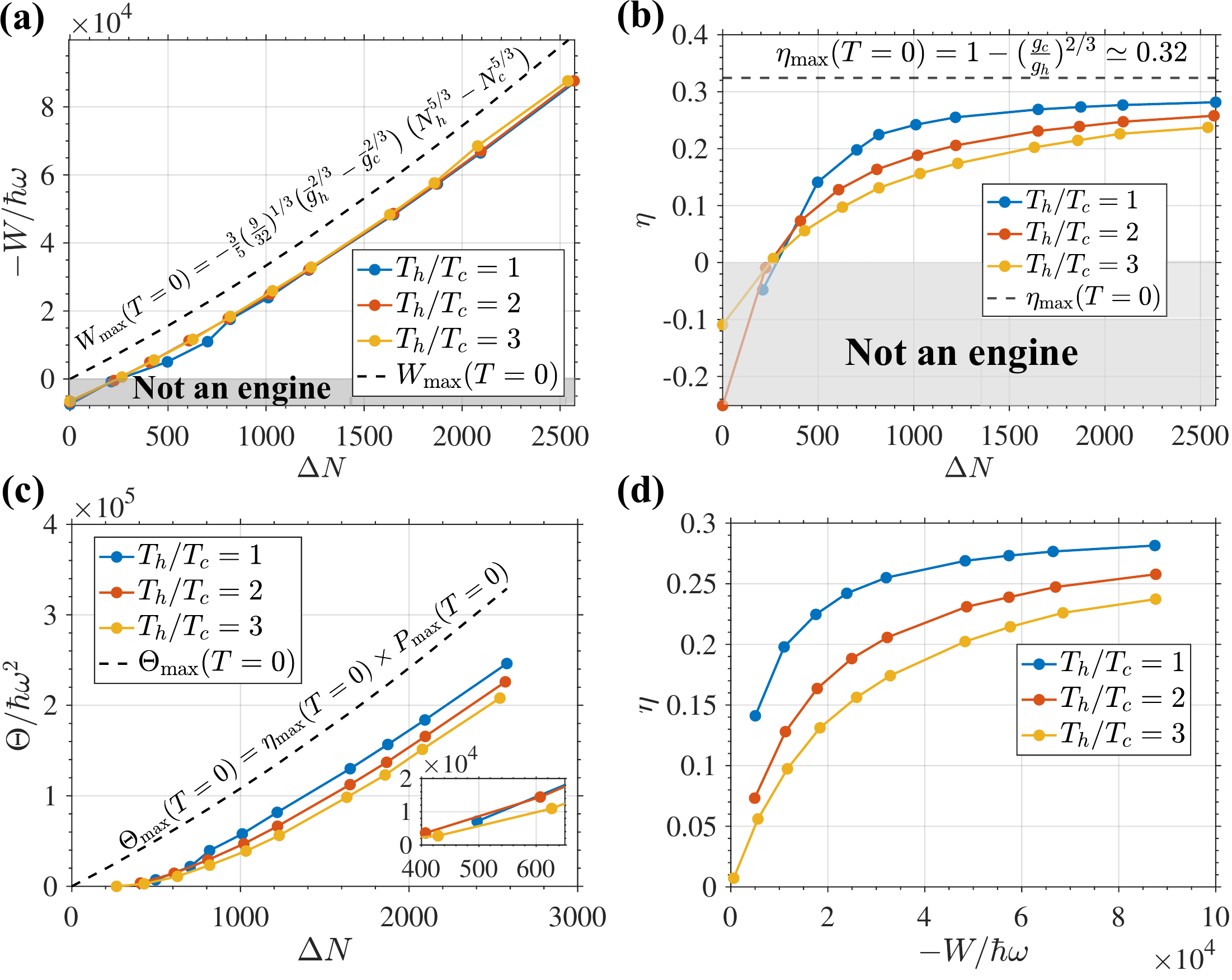}
    \caption{(a) Net work, $-W$; (b) efficiency, $\eta$; and (c) efficient power, $\Theta$; as functions of the number of particles, $\Delta N$, exchanged with the reservoir in the interaction-driven \textit{out-of-equilibrium} (sudden quench) QTE. Chemical work is performed on the working fluid via the inflow of $\Delta N$ particles from the hot reservoir. The results are plotted for three different temperature ratios between hot and cold reservoirs, $T_h/T_c$. In (d), efficiency is plotted as a function of the net beneficial work done by the QTE in one cycle. In (a), the shaded region indicates conditions where engine operation is not feasible, i.e., $-W < 0$. Data points with $-W < 0$ are excluded from panels (c) and (d). The data points in (d) represent the same range of $\Delta N$ as shown in (b) and (c). All other parameters are the same as in Fig.~\ref{fig:Fig2}.}
    \label{fig:Fig3}
\end{figure*}

\subsection{Effect of chemical work}
\label{deltaNeffect}

In Fig.~\ref{fig:Fig3}, we examine how chemical work, performed via the inflow of particles, affects the QTE's performance. Specifically, we investigate the impact of the number of particles, $\Delta N$, exchanged with the reservoir on key performance metrics of the QTE, such as net work, efficiency, and efficient power, while operating in the \textit{out-of-equilibrium} (sudden quench) regime. The performance is compared for three different temperatures of the hot reservoir, $T_h$, while keeping the temperature of the cold reservoir, $T_c$, fixed.

In Fig.~\ref{fig:Fig3}(a), the net work produced by the sudden quench QTE is plotted as a function of $\Delta N$ for the three values of $T_h$. The results reveal that engine operation is only enabled beyond a threshold value of $\Delta N \simeq 200$, indicating a minimum amount of chemical work is required to sustain operation as a QTE. As $\Delta N$ increases, the net beneficial work produced by the engine also increases. This occurs because performing additional chemical work on the working fluid increases the input energy, $E_\mathrm{in}$, transferred to the system during hot thermalization stroke \textbf{AB}, thereby supplying more energy for conversion into useful work during the subsequent expansion stroke \textbf{BC}.

Interestingly, we find that the net work is largely insensitive to the temperature ratio, $T_h/T_c$, suggesting that the temperature of the hot reservoir is not a critical parameter for optimizing the QTE. This observation aligns with recent findings in quantum engines using Bose gases \cite{estrada2024quantum}, which concluded that temperature plays a relatively minor role in improving engine performance for this working fluid.

Finally, as shown in Fig.\ref{fig:Fig3}(a), the zero-temperature QTE studied in Ref.\cite{keller2020feshbach} establishes an upper bound (black dashed line) on the net work produced by the QTE operating at finite (non-zero) temperatures. This upper bound is derived from the analytical expression for maximum work at zero temperature, given in Eq.(\ref{eq:maxwork}). These findings align with those in Fig.\ref{fig:Fig2}(a), where the zero temperature limit sets an upper bound on the maximum work of the QTE operating in a quasistatic regime.

Further, in Fig.~\ref{fig:Fig3}(a), the net work for the finite-temperature QTE is lower than in the zero-temperature case due to operation in the sudden quench regime, where irreversible work is generated by non-adiabatic excitations. Nevertheless, the difference is relatively minor, indicating that the amount of irreversible work produced in the sudden quench QTE will not reduce the efficiency by a significant amount.

\begin{figure*}[t!]
    \centering
    \includegraphics[scale=0.40]{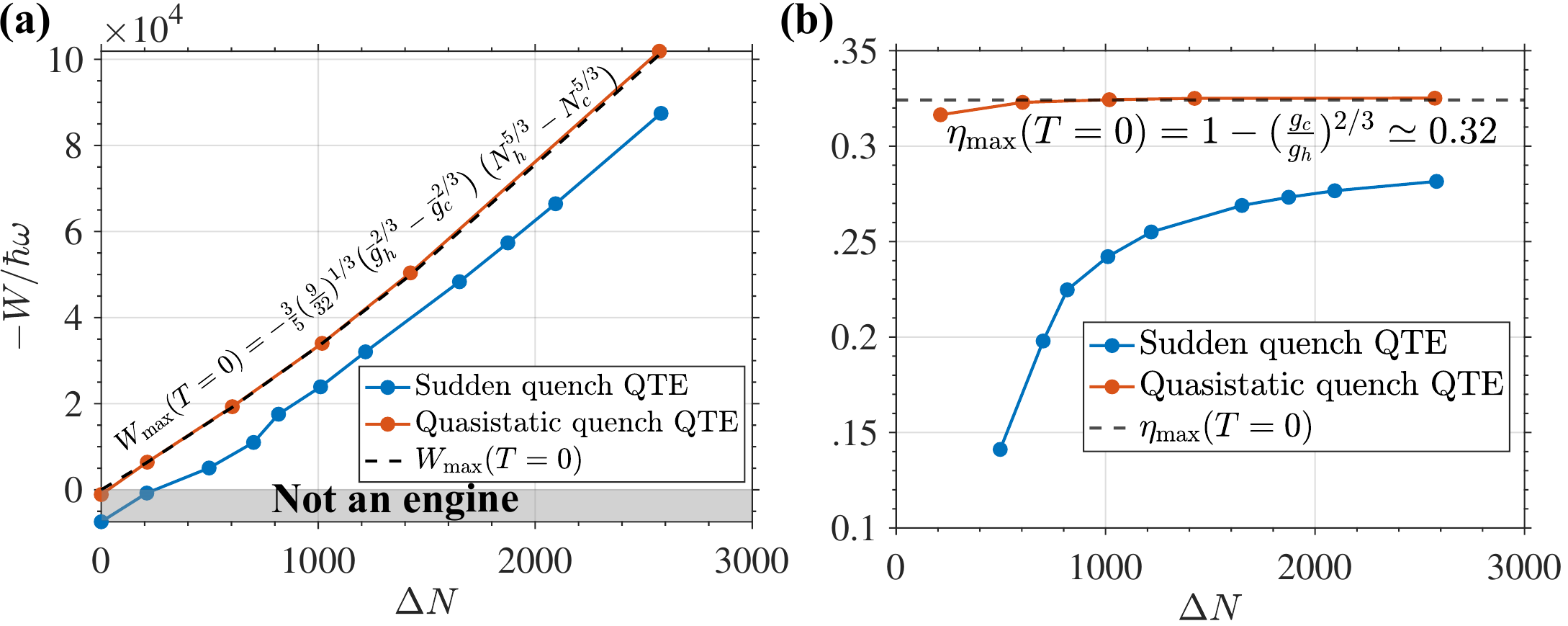}
    \caption{(a) Net work, $-W$ and (b) efficiency, $\eta$ as functions of the number of particles, $\Delta N$, exchanged with the reservoir in the interaction-driven \textit{out-of-equilibrium} (sudden quench) QTE (blue curve) and quasistatic QTE  (red curve). Chemical work is performed on the working fluid via the inflow of $\Delta N$ particles from the hot reservoir. In (a), the shaded region indicates conditions where engine operation is not feasible, i.e., $-W < 0$. Data points with $-W < 0$ are excluded from panel (b). All other parameters are the same as in Fig.~\ref{fig:Fig2}. The results in this figure are for $T_h/T_c=1$.}
    \label{fig:Fig4}
\end{figure*}

Next, in Fig.~\ref{fig:Fig3}(b), we plot the efficiency of the \textit{out-of-equilibrium} QTE as a function of $\Delta N$. Similar to net work, efficiency increases with $\Delta N$. However, unlike net work, we observe a saturation in efficiency once a sufficient amount of chemical work is performed on the working fluid. Beyond a certain threshold, $\Delta N \simeq 1300$, the efficiency saturates, showing no further significant increase. This is one of our main results because, as observed in Fig.\ref{fig:Fig:tq}, we identified a specific work stroke duration, $t_\mathrm{w} \simeq 10/\omega$, at which efficiency approaches near-maximum value, $\eta \rightarrow  \eta_\mathrm{max}(T=0)$ (see Eq.~\eqref{eq:max_efficiency}), but power output drops to zero, i.e., $P \rightarrow 0$ due to long duration of $t_\mathrm{w}$. Now, the results in Fig.~\ref{fig:Fig3}(b) demonstrate that for sufficiently large values of $\Delta N$, near-maximum efficiency can be achieved, $\eta \rightarrow \eta_\mathrm{max} (T=0)$. This suggests that instead of increasing $t_\mathrm{w}$ to enhance efficiency, a similar outcome can be attained by increasing $\Delta N$. This approach offers a favorable trade-off between power and efficiency, as maintaining $t_\mathrm{w}$ in the sudden quench regime reduces the engine's cycle time—enhancing overall power output—while simultaneously improving efficiency through the increase of $\Delta N$.

At high values of $\Delta N$, efficiency appears largely insensitive to temperature, similar to the behaviour of the net work. That said, a slight decrease in efficiency is observed as the temperature of the hot reservoir increases, particularly at lower values of $\Delta N$ ($\Delta N < 1000$). For instance, when the hot and cold reservoirs have the same temperature (blue curve), efficiency is highest, whereas for the hottest temperature of the reservoir (yellow curve), efficiency is lowest. These findings are consistent with the results in Fig.~\ref{fig:Fig2}(b), where efficiency decreases as the temperature of the hot reservoir increases. Notably, this gap narrows as $\Delta N$ increases, and for the largest value plotted ($\Delta N \simeq 2580$), efficiencies across different temperatures converge and are nearly identical.

Another important observation from Fig~\ref{fig:Fig3}(b) is that the adiabatic efficiency at zero temperature, $\eta_\mathrm{max}(T=0)$, given by Eq.\ref{eq:max_efficiency} for a zero-temperature QTE studied in Ref.\cite{keller2020feshbach}, serves as an upper bound for the efficiency of our proposed finite-temperature QTE. Additionally, for large values of $\Delta N$, the efficiency of the sudden quench (out-of-equilibrium) QTE approaches this upper limit, indicating a favourable trade-off between efficiency and power output. Notably, efficiency close to the maximum efficiency at zero temperature ($\eta_\mathrm{max}(T=0)$) is achieved even in the sudden quench regime—which typically yields the lowest efficiency—through the application of chemical work via particle inflow from the hot reservoir.

In Fig.~\ref{fig:Fig3}(c), we plot efficient power, $\theta$, as a function of $\Delta N$. As $\Delta N$ increases, the efficient power also rises, which is expected since an increase in $\Delta N$ leads to higher net work, thereby boosting efficient power. Similar to net work, the variation in $\theta$ appears largely insensitive to temperature changes, as all three temperature cases produce approximately similar values of efficient power.

In Fig.\ref{fig:Fig3}(d), we plot the variation of efficiency, $\eta$, with net work, $-W$. As expected, an increase in net work does not lead to a corresponding increase in efficiency, consistent with the trends observed in Fig.\ref{fig:Fig3}(a) and Fig.~\ref{fig:Fig3}(b).

\subsection{The zero temperature adiabatic QTE sets an upper bound on QTE in finite (non-zero) temperature operational regime}

In Fig.~\ref{fig:Fig4}, we compare the performance of the quasistatic QTE with that of the out-of-equilibrium (sudden quench) QTE. Specifically, we analyze (a) net work and (b) efficiency as functions of the number of particles, $\Delta N$, exchanged with the reservoir in the two cases.

In Fig.\ref{fig:Fig4}(a), we observe that the net work produced by the sudden quench QTE (blue curve) is close to that of the quasistatic QTE (red curve). Furthermore, we find that the near-maximum work of the finite-temperature QTE, obtained via the quasistatic near-adiabatic quench, matches the maximum work of the zero-temperature adiabatic QTE (black dashed curve), as predicted by Eq.(\ref{eq:maxwork}). The zero-temperature QTE operating in the adiabatic limit, as studied in reference \cite{keller2020feshbach}, sets an upper limit on the work output of the non-zero (finite) temperature QTE studied here. There is excellent agreement between the numerical results for finite-time operation in the quasistatic regime (red curve) and the analytical expression for the maximum work of the zero-temperature QTE (black-dashed curve). These findings are consistent with the results presented in Fig.~\ref{fig:Fig2}(a).

In Fig.~\ref{fig:Fig4}(b), we compare the efficiency of the sudden quench and quasistatic QTEs. Similar to the case of net work, the efficiency of the sudden quench engine (blue curve) is close to the near-maximum efficiency achieved by the quasistatic engine operating in near-adiabatic regime (red curve). However, achieving near-maximum efficiency in the sudden quench QTE requires a finite amount of chemical work, represented by the number of particles exchanged with the reservoir, $\Delta N$. For $\Delta N < 1000$, the efficiency gap between the sudden quench and quasistatic QTEs is relatively larger. As $\Delta N$ increases, or as more chemical work is performed through particle exchange with the reservoir, the sudden quench QTE (blue curve) efficiency approaches the quasistatic engine's maximum efficiency.

Additionally, in the quasistatic limit, as $\Delta N$ increases, the efficiency converges to the zero-temperature case studied in Ref.\cite{keller2020feshbach}. This observation confirms that the zero-temperature QTE establishes an upper bound on our proposed QTE's efficiency and work output in the finite (non-zero)-temperature regime. The results presented in Fig.\ref{fig:Fig4} are consistent with those shown in Fig.\ref{fig:Fig2}, Fig.\ref{fig:Fig3}, and Fig.\ref{fig:Fig5}. Across all these results, the zero-temperature adiabatic QTE \cite{keller2020feshbach} (black dashed line) consistently serves as the upper bound on both efficiency and net work for our finite-temperature QTE. Notably, in our finite-temperature QTE, this near-maximum efficiency is achieved even in the out-of-equilibrium regime, without employing any optimization protocols, simply by increasing the amount of chemical work ($\Delta N$), as demonstrated in Fig.\ref{fig:Fig3}(b) and Fig.\ref{fig:Fig4}(b).

\subsection{Effect of temperature on engine performance}
\label{temeffect}
Our numerical analysis reveals that the efficiency of the QTE is negatively affected by an increase in reservoir temperature. For example, in Fig.\ref{fig:Fig2}(b), we observe a decrease in efficiency as the reservoir temperature, $T_h$, increases. Similar trends can be inferred from the results presented in Fig.\ref{fig:Fig3}(b) and Fig.~\ref{fig:Fig5}. Unlike conventional Otto or Carnot heat engines, where efficiency typically depends directly on the temperature ratio of the hot and cold reservoirs, $T_h/T_c$ \cite{schroeder2020introduction}, the efficiency of the out-of-equilibrium QTE studied here decreases with an increase in heat input or reservoir temperature, and hence the ratio $T_h/T_c$.

The primary reason for this behaviour is that increasing the reservoir temperature raises the input energy, $E_\mathrm{in}$. This additional energy, supplied as heat, increases the operational cost without contributing to the production of beneficial work, as shown in Fig.~\ref{fig:Fig2}(a) (yellow and purple curves). Consequently, the overall efficiency decreases since efficiency is defined as $\eta = -W/E_\mathrm{in}$.

\subsection{Effect of compression ratio}
\label{ratioeffect}
  
\begin{figure}[t!]
    \centering
    \includegraphics[scale=0.40
    ]{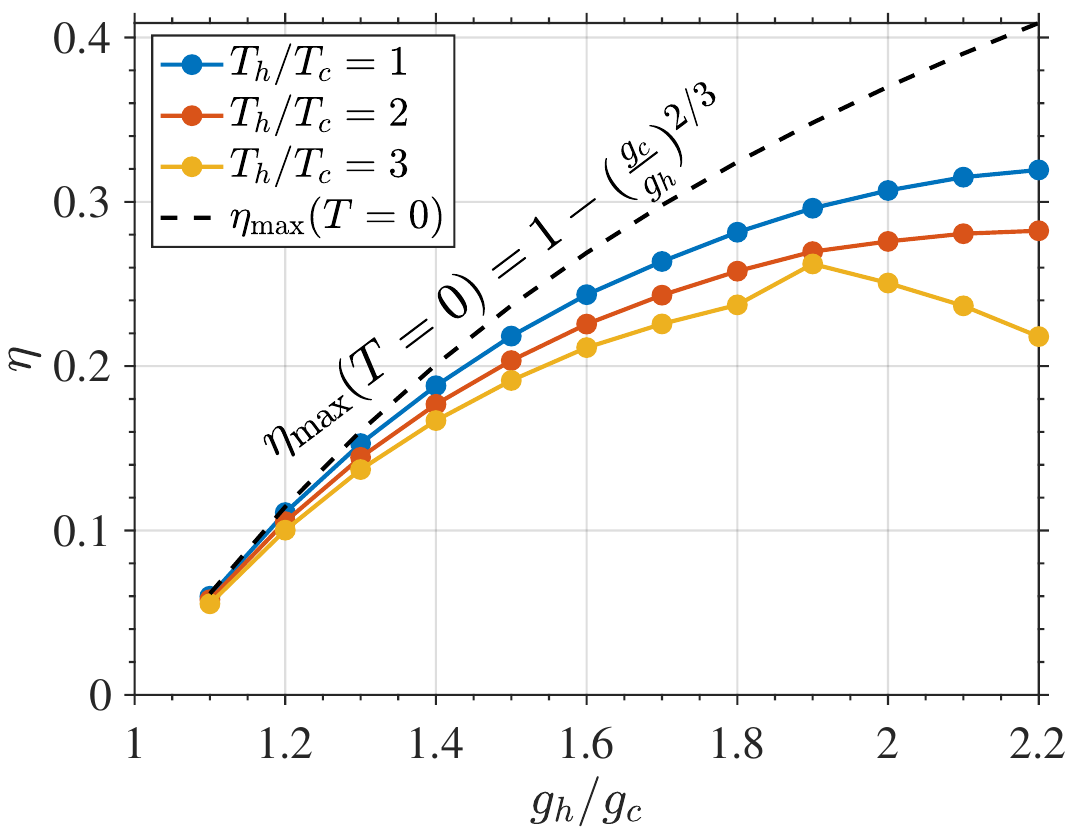}
    \caption{Efficiency, $\eta$ as a function of the compression ratio, $g_h/g_c$, in the \textit{out-of-equilibrium} (sudden quench) QTE. We plot the efficiency of sudden quench QTE for three different ratios of hot and cold reservoir temperatures, $T_h/T_c$. All other parameters are the same as in Fig.~\ref{fig:Fig2}.}
     \label{fig:Fig5}
\end{figure}

In Fig.\ref{fig:Fig5}, we present the variation in the efficiency of the out-of-equilibrium (sudden quench) QTE with the compression ratio, $g_h/g_c$, for different temperature ratios between the hot and cold reservoirs, $T_h/T_c$. The temperature $T_c$ and the interaction strength $g_c$ at the cold equilibrium point \textbf{D} (Fig.\ref{fig:Fig1}) were fixed, while the interaction strength $g_h$ and temperature $T_h$ at the hot equilibrium point \textbf{A} were varied.

At small values of $g_h/g_c$, the efficiency is low, and the finite-temperature efficiencies (colored curves) closely match the zero-temperature efficiency, $\eta_\mathrm{max}(T=0)$, represented by the black dashed curve and given by Eq.~\ref{eq:max_efficiency}. However, as the compression ratio increases, the efficiency also increases, though the difference between the zero-temperature case and finite-temperature efficiencies becomes more pronounced compared to lower values of $g_h/g_c$. Consistent with previous results, an increase in the reservoir temperature also reduces the efficiency in this scenario. The highest reservoir temperature (yellow curve) yields a lower efficiency than the lowest temperature (blue curve). Once again, the zero-temperature adiabatic QTE was studied in reference.~\cite{keller2020feshbach} establishes an upper bound for the efficiency of our finite-temperature QTE.

\section{Conclusions} \label{sec:conclusion}

We theoretically studied an interaction-driven quantum thermochemical engine (QTE) operating in an Otto cycle utilizing a harmonically trapped, weakly interacting 1D Bose gas as the working fluid. During the interaction-driven work strokes, the working fluid is treated as an isolated quantum many-body system undergoing dynamic evolution beginning from an initial thermal equilibrium state. To implement the thermalization strokes, the working fluid was modelled as an open quantum many-body system in diffusive contact with a thermal reservoir. This diffusive contact allowed us to perform additional chemical work (in addition to heat) on the working fluid through particle exchange with the reservoir, enabling operation as a QTE.

We evaluated the finite-time performance of the QTE and quantified key figures of merit, focusing on its operation in the \textit{out-of-equilibrium} (sudden quench) regime. In this regime, work strokes are implemented in the shortest possible time, driving the working fluid out of equilibrium. This presents an opportunity to enhance the engine's power output due to fast driving.

The performance of the \textit{out-of-equilibrium} QTE was benchmarked against two scenarios: (\emph{i}) an adiabatic (maximum-efficiency) QTE operating at zero temperature, as studied in reference ~\cite{keller2020feshbach}; and (\emph{ii}) a quasistatic engine operating at realistic experimental temperatures, where work strokes are completed extremely slowly to achieve near-maximum finite-time efficiency.

Our findings show that the zero-temperature QTE (studied in reference \cite{keller2020feshbach}) establishes an upper bound on the work and efficiency achievable by the proposed QTE operating with a finite-temperature Bose gas explored in this work. Additionally, we explored the trade-off between efficiency and power output in the \textit{out-of-equilibrium} (sudden quench) regime of operation. Remarkably, we found that the proposed QTE can operate at near-maximum efficiencies even in the out-of-equilibrium (sudden quench) regime while maintaining high power output. This high efficiency is achieved through the chemical work performed on the working fluid via particle inflow from the hot reservoir. Notably, efficiencies close to the quasistatic (near-maximum) limit were attained using a simple finite-time sudden quench protocol without employing any optimization techniques such as the shortcut to adiabaticity used in reference ~\cite{keller2020feshbach}.

We emphasize, however, that the efficiency of our proposed QTE will always remain lower than that of an engine operating under similar conditions but utilizing optimization protocols, such as the STA, instead of a simple linear sudden quench protocol explored here. This reduction in efficiency arises due to quantum friction \cite{ccakmak2016irreversibility,campbell2017trade,abah2019shortcut} during the interaction-driven work strokes generating irreversible work. In a 1D Bose gas, this irreversible work manifests in the form of breathing mode excitations (e.g., as shown in Refs.~\cite{bouchoule2016finite, bayocboc2023frequency, nautiyal2024finite, tschischik2013breathing, fang2014quench}), which arise due to a quench in interaction strength (as in this work) or a quench in trapping frequency (as in a conventional volumetric Otto cycle \cite{estrada2024quantum}) used to implement the work strokes. Thus, the efficiency of the proposed \textit{out-of-equilibrium} QTE is always lower than the quasistatic efficiencies or the zero-temperature adiabatic efficiencies achieved via STA \cite{keller2020feshbach}. Nonetheless, our analysis reveals that coupling with a thermochemical reservoir allows for mitigating these losses due to quantum friction to a great extent by utilizing additional chemical work. This allows for avoiding practical challenges associated with STA protocols, such as modulation instability and additional energy cost for implementing the optimization protocols \cite{keller2020feshbach, li2018efficient, calzetta2018not}.

This study provides a foundation for exploring the performance of the proposed QTE across a broader parameter space of 1D Bose gases. In Ref. \cite{marzolino2024quantum}, a conventional (volumetric) Otto QTE with an ideal (non-interacting) homogeneous quantum gas confined in a box potential was investigated. It was found that when the working fluid was a quantum gas—whether Bose or Fermionic—the QTE outperformed a similar engine operating with a classical gas. Likewise, in Ref.\cite{simmons2023thermodynamic}, an Otto-like engine with all four unitary strokes was studied. A significant enhancement in both power and efficiency was observed when the working fluid was a degenerate quantum Bose gas compared to a non-degenerate classical gas.

A similar comparative analysis could extend the work presented in this paper by examining the performance of a QTE with a weakly interacting harmonically trapped Bose gas (as studied here) in comparison to a QTE with a 1D Bose gas in the classical or non-degenerate regime (see references \cite{kheruntsyan2003pair,kheruntsyan2005finite} for details on various physical regimes available in 1D Bose gases). Such a comparison could provide insights into whether quantum degeneracy offers a performance advantage in QTE operation. However, the $c$-field numerical approach used in this study is not valid in the classical or strongly interacting regimes of 1D Bose gases (see Appendix B of Ref.~\cite{bayocboc2023frequency} for a detailed discussion on the validity of the $c$-field approach).

While our analysis focuses on the weakly interacting (quasicondensate) regime, 1D Bose gases exhibit a rich parameter space, first identified in Refs.~\cite{kheruntsyan2003pair, kheruntsyan2005finite}. A natural extension of this work would be to investigate the performance of the out-of-equilibrium QTE in the strongly interacting Tonks-Girardeau regime. Since the $c$-field method used here is best suited for weakly interacting gases, future studies could employ the theory of generalized hydrodynamics \cite{malvania2021generalized,bouchoule2022generalized, watson2024quantumthermodynamicsintegrablenearintegrable} to comprehensively evaluate and compare the QTE’s performance across a wider range of interaction strengths in 1D Bose gases.

\section*{Acknowledgements}

The author acknowledges support from the Australian Government Research
Training Program Scholarship.

\section*{Appendices}

In these appendices, we describe the differences between an ideal \textit{heat} engine, such as the Carnot engine, and our proposed quantum thermal engine (QTE). Specifically, in \textbf{Appendix~A}, we show how, due to fundamental differences in the operational scheme, the Carnot efficiency is not a strict upper bound for a QTE, unlike for a traditional \textit{heat} engine. Next, in \textbf{Appendix~B}, we describe the physical processes involved during the thermalization stroke of the QTE. Specifically, we detail the grand-canonical coupling between the working fluid and the thermal reservoir that facilitates the exchange of particles.

 \section*{Appendix A: Validity of the laws of thermodynamics in the proposed QTE-- Comparison with Carnot efficiency}
 \label{appendexA}

\subsubsection{Maximum efficiency of a \textit{heat} engine}

\begin{figure}[t!]
    \centering
    \includegraphics[scale=0.40
    ]{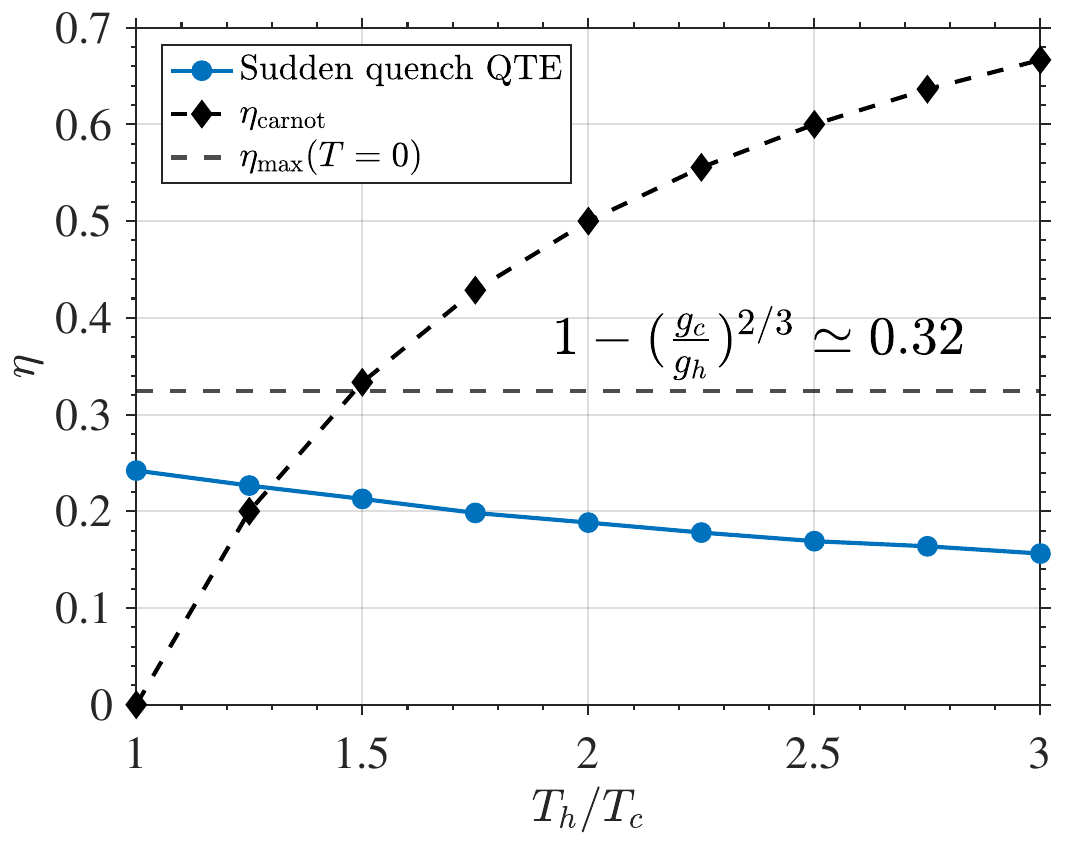}
    \caption{Efficiency, $\eta$, as a function of the temperature ratio of the hot and cold reservoirs, $T_h/T_c$, for a sudden quench QTE (blue curve) and a Carnot engine (black dashed diamond curve). The blue curve and all parameters remain the same as in Fig.~\ref{fig:Fig2}(b), while the black dashed diamond curve represents the Carnot efficiency for comparison.}
     \label{fig:Apendix2}
\end{figure}

In Fig.~\ref{fig:Apendix2}, we compare the efficiency of our proposed sudden quench QTE with that of a Carnot engine operating between the same temperature ratios of the thermal reservoirs. At low temperatures, the efficiency of our QTE appears to surpass the Carnot limit, which may suggest a violation of classical thermodynamic bounds. However, as we clarify below, our engine operates fully within the principles of thermodynamics. We explain why the Carnot efficiency is not the appropriate upper bound for our QTE and how the additional chemical work modifies the thermodynamic constraints.

In general, for any thermodynamic cycle, four thermodynamically allowed modes of operation exist \cite{watson2024interaction, buffoni2019quantum, bhattacharjee2021quantum}: operation as an engine, refrigerator, accelerator, or heater (heat pump). The performance of these quantum thermal machines (QTMs) is characterized by their coefficient of performance (CoP), which is given by the general relation \cite{schroeder2020introduction, watson2024interaction}:

\begin{equation} \label{eqn:cop} CoP[QTM] = \frac{\mathrm{benefit\ of\ operation}}{\mathrm{cost\ of\ operation}}. \end{equation}

For an engine, the CoP is generally referred to as the efficiency, $\eta$. The ``benefit'' of operation is the net work output of the working fluid, while the ``cost'' of operation is the energy supplied to the working fluid from the hot reservoir. From the first law of thermodynamics (energy conservation), we have:

\begin{equation} W_\mathrm{com} + W_\mathrm{exp} = E_\mathrm{in} + E_\mathrm{out}. \end{equation}

Here, net beneficial work is obtained when $|W_\mathrm{exp}| > W_\mathrm{com}$, or equivalently, when $E_\mathrm{in} > |E_\mathrm{out}|$. Substituting this into Eq.~(\ref{eqn:cop}), we obtain a general expression for the engine efficiency:

\begin{equation} 
\label{eqn:etageneral}
\eta = -\frac{W_\mathrm{com} + W_\mathrm{exp}}{E_\mathrm{in}} = 1 - \frac{|E_\mathrm{out}|}{E_\mathrm{in}}. \end{equation}

Now, for a conventional heat engine, such as the standard Otto cycle, the working fluid exchanges only heat with the reservoirs and follows a canonical ensemble description. In this case, the energy supplied to the working fluid consists purely of heat, i.e., $E_\mathrm{in} = Q_h$ and $E_\mathrm{out} = Q_c$, where $Q_h$ and $Q_c$ represent:

\begin{itemize}
    \item $Q_h$: Heat absorbed by the working fluid from the hot reservoir during isochoric thermalization with the hot reservoir at temperature $T_h$.
    \item $Q_c$: Remaining heat expelled by the working fluid into the cold reservoir during isochoric thermalization with the cold reservoir at temperature $T_c$.
\end{itemize}
    
Therefore, the efficiency of a heat engine is given by \cite{schroeder2020introduction}:

\begin{equation} \eta = 1 - \frac{|Q_c|}{Q_h}. \end{equation}
Further, from the second law of thermodynamics (total entropy of the working fluid and the reservoirs will always increase), we obtain \cite{schroeder2020introduction}
\begin{equation}
    \frac{|Q_c|}{Q_h} \geq \frac{T_c}{T_h}.
\end{equation}

Carnot \cite{carnot1978reflexions} showed that for an ideal irreversible engine cycle, $\frac{|Q_c|}{Q_h} = \frac{T_c}{T_h}$. Therefore, by combining the first and second law of thermodynamics \cite{schroeder2020introduction, verley2014unlikely, lucia2013carnot,bender2000quantum}, we obtain the maximum efficiency of a \textit{heat} engine operating between two thermal reservoirs at temperatures $T_c$ and $T_h$, known as the Carnot efficiency \cite{carnot1978reflexions,schroeder2020introduction},
\begin{equation}
\label{eqn:carnot}
    \eta_\mathrm{carnot} = 1 - \frac{T_c}{T_h}.
\end{equation}

\subsubsection{Why the Carnot efficiency is not the upper bound for the proposed \textit{thermochemical} engine}
Our proposed QTE is fundamentally different from a conventional \textit{heat} engine. We underscore that Eq.~\eqref{eqn:carnot} is derived assuming canonical ensemble between the reservoir and the working fluid, i.e., if during the thermalization strokes, the increase in internal energy of the working fluid is solely due to heat absorbed from the reservoir ($E_\mathrm{in}=Q_h$). Unlike a standard Otto cycle, our QTE operates under a grand-canonical ensemble description, where the working fluid exchanges both heat and particles with the thermal reservoirs.

 \begin{figure}[t!]
    \centering
    \includegraphics[scale=0.25]{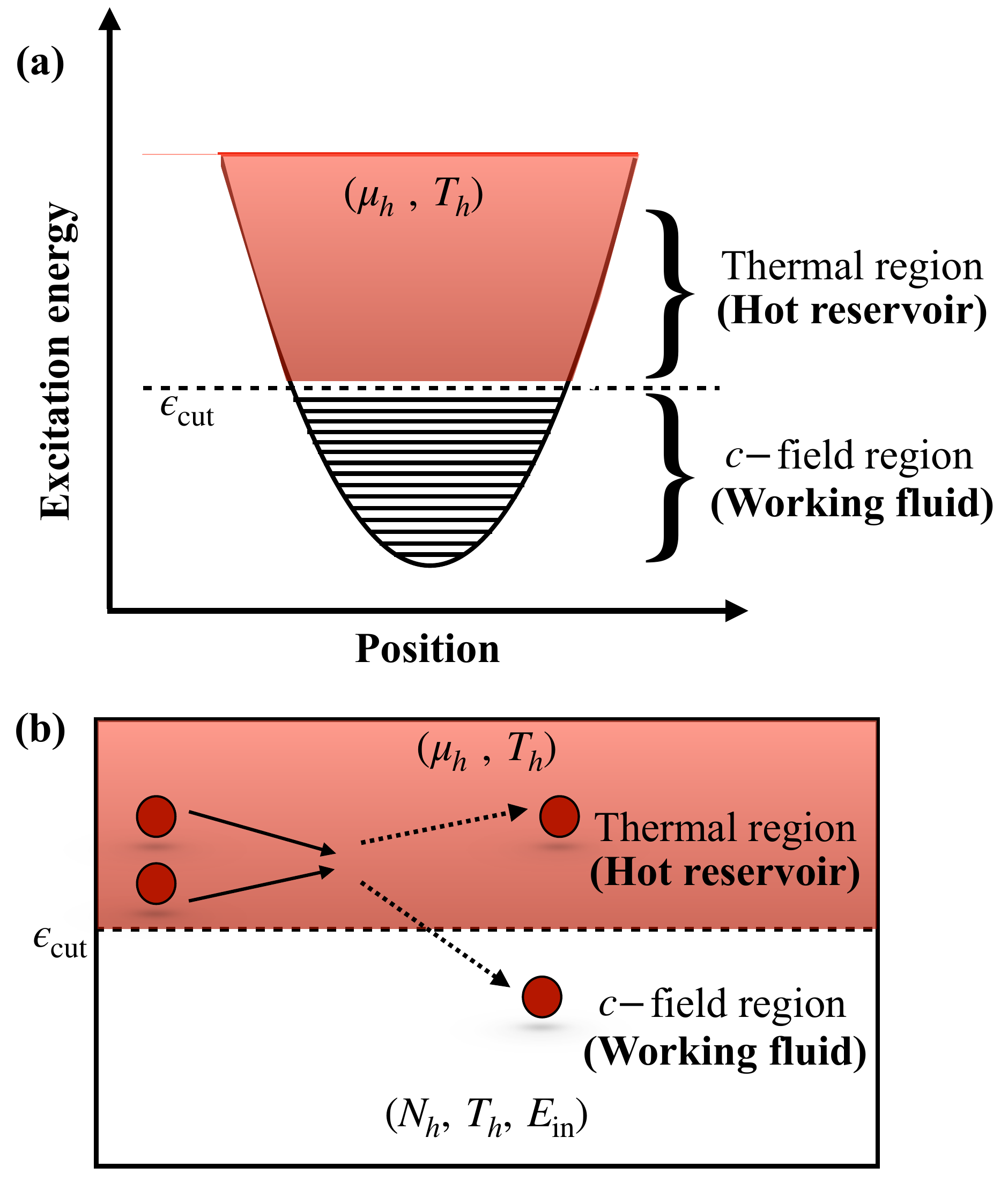}
    \caption{Schematic representation of physical processes involved in thermalization stroke \textbf{AB} with the hot reservoir.(a) The thermal region (hot reservoir) and the $c$-field region (working fluid) (see Sec.\ref{subsec:numericmethod} in the main text) are separated by the energy cut-off, $\epsilon_\mathrm{cut}$, which is determined by the projection operator, $\mathcal{P}^{\mathbf{C}}$ (see Eqs.~\eqref{eq:SPGPE} and \eqref{eq:spgpe_hot}). The $c$-field region consists of highly occupied low-energy modes, while the thermal region contains sparsely occupied high-energy modes \cite{rooney2012stochastic, rooney2013persistent, rooney2010decay}. The temperature, $T_h$, and the chemical potential, $\mu_h$, of the thermal region set the temperature and particle number in the $c$-field (working fluid). (b) Working fluid–reservoir interaction processes during the thermalization strokes. In the process illustrated, two high-energy atoms in the thermal region (hot reservoir) collide, causing one atom to transfer into the $c$-field (working fluid) while the other remains in the thermal region. This type of collision, described by the grand-canonical ensemble, is often referred to as the ``growth'' process because it alters both the particle number and energy of the $c$-field region. Consequently, at the end of the thermalization stroke with the hot reservoir, the working fluid (or the $c$-field region) contains $N_h$ particles and reaches thermal equilibrium with the thermal region (hot reservoir) at temperature $T_h$. A similar description applies to the thermalization stroke \textbf{CD} with the cold reservoir, except that in this case, excess particles, $\Delta N$, transfer from the $ c$ field to the thermal bath. The chemical potential and temperature of the thermal region are adjusted so that, at the end of the thermalization stroke with the cold reservoir, the $c$-field contains $N_c=N_h-\Delta N$ particles and achieves a temperature, $T_c$ in equilibrium with cold thermal reservoir.}  
     \label{fig:Apendix3}
\end{figure}
 The total input energy (or the \textit{cost} of operation), $E_\mathrm{in}$ in our QTE has contributions from both heat transfer and chemical work components, i.e.,
\begin{equation}
\label{eq:einQTE}
    E_\mathrm{in} = Q_h + W_\mathrm{chem},
\end{equation}
where $W_\mathrm{chem}$ is the additional chemical work done on the working fluid through the inflow of $\Delta N$ particles from the hot reservoir.

Since, for our proposed QTE, $E_\mathrm{in}$ is given by Eq.~\eqref{eq:einQTE} (instead of simply $E_\mathrm{in}=Q_h$), therefore the Carnot efficiency, $\eta_\mathrm{carnot}$ given in Eq.~\eqref{eqn:carnot} is no longer an appropriate upper bound for the QTE. Instead, the efficiency of the QTE is then given by general expression for the efficiency (eq.~\eqref{eqn:etageneral}), i.e.
\begin{equation} \eta_{\mathrm{QTE}} = 1 - \frac{|E_\mathrm{out}|}{E_\mathrm{in}} = 1 - \frac{|Q_c| + |W^{'}_\mathrm{chem}|}{Q_h + W_\mathrm{chem}}, \end{equation}
where $W^{'}_{\mathrm{chem}}$ represents the unused chemical energy expelled by the working fluid (through the outflow of $\Delta N$ particles-- taken from the hot reservoir-- into the cold reservoir), along with the unused heat, $Q_c$, to the cold reservoir, at the end of thermalization stroke \textbf{CD}. This returns the working fluid to the initial state at the start of the Otto cycle at point \textbf{D}.

In theory, the second law of thermodynamics does not limit the maximum efficiency of a quantum thermal engine (QTE), and it does not rule out the possibility of converting all the supplied chemical work into mechanical work without any waste \cite{marzolino2024quantum}. For instance, reference \cite{marzolino2024quantum} demonstrates that, under certain conditions, in an ideal, reversible scenario, a QTE with a degenerate, non-interacting quantum gas in a box potential can achieve a maximum efficiency of one, i.e., $\eta_\mathrm{rev} \rightarrow 1$.

Since, in ultracold atomic gases, heat and particle transports are intrinsically coupled processes (for example, see references \cite{husmann2018breakdown, brantut2013thermoelectric}), therefore, we cannot precisely calculate individual contributions, $Q_{h(c)}$ and $W_\mathrm{chem}$, in the total energy $E_{\mathrm{in}(\mathrm{out})}$. However, we emphasize that the chemical work is accounted for in the total energetic cost when evaluating the efficiency of the Otto \textit{thermochemical} engine. This is done by defining efficiency as
$\eta = -W/E_\mathrm{in}$,
rather than using the conventional heat-engine definition, $\eta = -W/Q_h$.

We numerically compute $E_\mathrm{in} = \langle \hat{H} \rangle_{\mathbf{B}}-\langle \hat{H} \rangle_{\mathbf{A}}>0$ and  $E_\mathrm{out} = \langle \hat{H} \rangle_{\mathbf{D}}-\langle \hat{H} \rangle_{\mathbf{C}}<0$ to calculate the efficiency $\eta$ for an engine operating with a finite-temperature Bose gas.  In both the quasistatic and sudden quench regimes, the efficiency and net work remains effectively bounded by the theoretical upper limit derived for a zero-temperature Bose gas, given by Eq.\eqref{eq:max_efficiency}. This bound was obtained analytically using the Thomas-Fermi approximation in Ref.\cite{keller2020feshbach}. The fact that our engine does not surpass this limit confirms that it operates within the fundamental constraints of thermodynamics and fully adheres to the laws of thermodynamics.

\section*{Appendix~B: The $c$-field approach to simulate the diffusive thermalization strokes with a reservoir-- A grand-canonical description of system-reservoir coupling}
\label{apendix:grandcanocnical}
We used the  $c$-field method \cite{Blakie_cfield_2008} to simulate the entire Otto cycle at finite temperatures. As shown in Fig.~\ref{fig:Apendix3}(a), the $c$-field theory divides the system into two distinct regions: \\

(\romannumeral 1) \textbf{The $c$-field region (working fluid)}: This region consists of highly occupied low-energy modes that can be described by a complex field amplitude, $\psi^{\mathbf{C}}$. The $c$-field region (see Fig.~\ref{fig:Apendix3}(a)) is the region of partially condensed matter waves \cite{rooney2016reservoir} that include the condensate as well as the low-energy excitations  \cite{rooney2016reservoir,rooney2010decay,blakie2005projected}. This $c$-field region serves as the working fluid in our proposed QTE.

(\romannumeral 2) \textbf{The thermal region (hot/cold reservoir)}: This region consists of sparsely occupied high-energy modes (shaded red in Fig.~\ref{fig:Apendix3}(a) and (b)). These high-energy modes are typically thermalized and act as an effective reservoir for the $c$-field (working fluid).

For simulating the isolated work strokes of the QTE, we use the Projected Gross-Pitaevskii Equation (PGPE) (Eq.~\eqref{eqn:PGPE}), which neglects the coupling between the $c$-field and the thermal reservoir. This describes the unitary evolution of the $c$-field \cite{blakie2005projected,rooney2016reservoir}. Since the PGPE conserves both the particle number and total energy of the system, it is well-suited for modeling the work strokes, effectively treating the $c$-field as a micro-canonical system.

In contrast, for simulating the thermalization strokes, where the working fluid is in diffusive contact with the thermal reservoir, we use the Stochastic Projected Gross-Pitaevskii Equation (SPGPE) (Eq.\eqref{eq:spgpe_hot}). The SPGPE includes additional terms that account for dissipative collision processes between high-energy atoms in the thermal region and the $c$-field \cite{weiler2008spontaneous}. These processes lead to both particle and energy exchange between the reservoir and the $c$-field, as depicted in Fig.\ref{fig:Apendix3}(b) \cite{rooney2016reservoir,rooney2012stochastic}. Experimentally, the dynamics of finite-temperature Bose gas, similar to those described via the thermalization stroke using the SPGPE, have been realised in previous studies (see references \cite{weiler2008spontaneous,rooney2010decay}).

Since both energy and particle exchange occur in the collision processes described by the SPGPE, the thermalization strokes in our proposed QTE correspond to a grand-canonical system-reservoir coupling. This type of description is often referred to as the \textit{simple growth} SPGPE \cite{bayocboc2022dynamics,rooney2012stochastic,rooney2016reservoir,blakie2008dynamics}. In principle, an alternative version of the SPGPE, known as the \textit{scattering} SPGPE \cite{rooney2012stochastic,rooney2014numerical}, could be used to model processes in which only energy transfer occurs without any associated particle exchange. However, since the thermalization strokes in our proposed QTE involve chemical work in the form of particle transfer, we employ the simple growth SPGPE, which accurately captures the working fluid–reservoir dynamics and ensures we obtain the correct final thermal state required for computing energy and efficiency \cite{rooney2012stochastic}.

\bibliographystyle{unsrt}
\bibliography{references}

\begin{thebibliography}{100}
\bibitem{parameters}
The initial temperature of the working fluid, in thermal equilibrium with the cold reservoir, was kept constant at $T_c = 86.3$ nK, with an initial particle count $N_c \simeq 2235$ obtained using a chemical potential of $\mu = 7.95 \times 10^{-31}$ J. In all cases, the interaction strength was quenched from $g_c = 1.27 \times 10^{-38}$ J $\cdot$ m to $g_h=1.8 \times g_c$ following Eq.(\ref{eq:g_com}); during expansion, it was quenched back to $g_c$ per Eq.(\ref{eq:g_exp}). All results in this work pertain to a gas of $^{87}$Rb atoms with mass $m \simeq 1.44 \times 10^{-25}$ kg, 3D scattering length $a_\mathrm{s} \simeq 5.31,$ nm, trap frequency $\omega/2\pi = 20.0$ Hz, and intial transverse confinement frequency $\omega_{\perp}/2\pi = 1.81$ kHz, achieving the interaction strength as $g_c \simeq 2 \hbar \omega_{\perp}a$. In this and all subsequent figures, the timescale for the sudden quench is $t_\mathrm{w} = 0.05/\omega$, while for the quasistatic quench, it is $t_\mathrm{w} = 300/\omega$.
\end{thebibliography}

\end{document}